\documentstyle[12pt]{article}

\def\be{\begin{equation}}
\def\ee{\end{equation}}
\def\bea{\begin{eqnarray}}
\def\eea{\end{eqnarray}}

\def\br{\begin{thebibliography}{abc}}
\def\er{\end{thebibliography}}
\def\rf{\bibitem}

\def\cstars{C$^*$-algebras }
\def\cstar{C$^*$-algebra }
\def\sdp{\hbox{ \raisebox{.25ex}{\tiny $|$}\hspace{.2ex}{$\!\!\times $}} }

\def\a{\alpha}
\def\b{\beta}
\def\c{\raisebox{.4ex}{$\chi$}}
\def\d{\delta}
\def\e{\epsilon}
\def\f{\phi}
\def\g{\gamma}
\def\h{\eta}

\def\l{\lambda}
\def\m{\mu}

\def\p{\pi}
\def\q{\theta}
\def\r{\rho}
\def\s{\sigma}

\def\x{\xi}

\def\D{\Delta}

\def\L{\Lambda}

\def\S{\Sigma}

\def\ca{{\cal A}}
\def\cb{{\cal B}}
\def\cc{{\cal C}}

\def\ce{{\cal E}}

\def\ch{{\cal H}}

\def\ck{{\cal K}}

\def\co{{\cal O}}
\def\cp{{\cal P}}

\def\ct{{\cal T}}
\def\cu{{\cal U}}

\def\rt{\rightarrow}

\def\pa{\partial}
\def\del{\nabla}
\def\iff{\Leftrightarrow}

\def\bar#1{\overline{#1}}

\def\Hat#1{\rlap{\kern.10em$\widehat{\phantom G}$}#1}
\def\HAt#1{\rlap{\kern.05em$\widehat{\phantom G}$}#1}

\def\czp#1{\rlap{\kern.1em$\widehat{\phantom{G\vrule height.8em}}$}#1{}}
\def\Czp#1{\rlap{\kern.05em$\widehat{\phantom{G\vrule height.8em}}$}#1{}}

\newcommand{\sect}[1]{\setcounter{equation}{0}\section{#1}}
\newcommand{\subsect}[1]{\subsection{#1}}

\def\fn{\footnote}
\def\sxn#1{\bigskip\medskip \sect{#1} \smallskip
                                                 }
\def\subsxn#1{\medskip \subsect{#1} \smallskip
                                                }

\begin{document}

\thispagestyle{empty}
\setcounter{page}{0}

\hfill ICTP: IC/94/38

\hfill Napoli:  DSF-T-2/94

\hfill Syracuse: SU-4240-567

\hfill March 1994

\vspace{.5cm}

\centerline {\LARGE FINITE QUANTUM PHYSICS AND}
\vspace{5mm}
\centerline {\LARGE NONCOMMUTATIVE GEOMETRY }
\vspace*{1.25cm}
\centerline {\large A.P. Balachandran$^1 $,
                    G. Bimonte$^{2,3}$,
                    E. Ercolessi$^1 $,
                    G. Landi$^{1,3,4,}$\fn[4]
{Fellow of the Italian National Council of Research (CNR) under Grant
No. 203.01.60.},}
\vspace{2.5mm}
\centerline{\large  F. Lizzi$^{3,5}$,
                    G. Sparano$^{3,5}$ and
                    P. Teotonio-Sobrinho$^1$}
\vspace{.75cm}
\centerline {\it $^1$ Department of Physics, Syracuse University,
Syracuse, NY 13244-1130, USA.}
\vspace{2.25mm}
\centerline {\it $^2$ International Centre for Theoretical Physics,
P.O. Box 586, I-34100, Trieste, Italy.}
\vspace{2.25mm}
\centerline {\it $^3$ INFN, Sezione di Napoli, Napoli, Italy.}
\vspace{2.25mm}
\centerline{\it $^4$ Dipartimento di Scienze Matematiche,
Universit\'a di Trieste,}
\centerline{\it P.le Europa 1, I-34127, Trieste, Italy.}
\vspace{2.25mm}
\centerline {\it $^5$ Dipartimento di Scienze Fisiche, Universit\`a di
Napoli,}
\centerline{\it Mostra d' Oltremare, Pad. 19, I-80125, Napoli, Italy.}

\vspace{.5cm}
\begin{abstract}
Conventional discrete approximations of a manifold do not preserve its
nontrivial topological features. In this article we describe an approximation
scheme due to Sorkin which reproduces physically important aspects of manifold
topology with striking fidelity. The approximating topological spaces in this
scheme are partially ordered sets (posets). Now, in ordinary quantum physics
on a manifold $M$, continuous probability densities generate the commutative
C*-algebra $\cc(M)$ of continuous functions on $M$. It has a fundamental
physical significance, containing the information to reconstruct the
topology of $M$, and serving to specify the domains of observables like the
Hamiltonian. For a poset, the role of this algebra is assumed by a
noncommutative C*-algebra $\ca $. As noncommutative geometries are based on
noncommutative C*-algebras, we therefore have a remarkable connection between
finite approximations to quantum physics and noncommutative geometries. Various
methods for doing quantum physics using $\ca $ are explored. Particular
attention is paid to developing numerically viable approximation schemes which
at the same time preserve important topological features of continuum physics.
\end{abstract}



\newpage
\setcounter{page}{1}

\sxn{Introduction}\label{se:1}

Experience teaches us that realistic physical theories are complicated and
require approximations for extraction of their predictions. A powerful
approximation method, particularly effective for numerical work, is the
discretisation of continuum physics where manifolds are replaced by a
lattice of points. It has acquired a central role in the study of
fundamental physical theories such as QCD \cite{DICK} or Einstein gravity
\cite{num}.

A notable limitation of such discretisations is their poor ability to
preserve the topological properties of continuum theories.
Thus in these approximations a manifold is typically substituted by a
 set of points with discrete topology. The latter is entirely incapable of
describing any significant topological attribute of the continuum, this
being equally the case for both local and global properties. There is for
example no nontrivial concept of winding number when manifolds are
modelled by discrete points and hence also no way to associate solitons
with winding numbers in these approximations.

Some time ago, Sorkin \cite{So} studied a very interesting method for
finite approximations of
manifolds by certain point sets in detail. [See also ref. \cite{Al}.]
These sets are partially ordered sets
(posets) and have the ability to reproduce important topological
features of the continuum with remarkable fidelity. Subsequent
research
\cite{BBET} developed these methods of Sorkin and others
and made them usable for
approximate computations in quantum physics. They could thus become
viable alternatives to computational schemes like those in lattice QCD
\cite{DICK}.

In this article, after a review of the poset approximation scheme in
Section \ref{se:2}
, we explore its properties in a novel direction. In quantum
physics on a manifold $M$, a fundamental role is played by the (C*-)
algebra $\cc(M)$ of continuous functions on $M$. Indeed, it is
possible to recover $M$, its topology and even its $C^\infty $-structure
when this algebra and a distinguished subalgebra are given
\cite{FD,Ma,Co,VG,CL}. For this reason, it is also possible to rewrite
quantum theories on $M$ substituting this algebra for $M$, the tools for doing
such calculations efficiently also being readily available
\cite{Co,VG,CL}.
All this material on $\cc(M)$ is reviewed in Section \ref{se:3}
 with particular
attention to its physical meaning.

The special role of $\cc(M)$ for manifolds suggests that it is of
basic interest to know the algebra $\ca $ replacing $\cc(M)$ when $M$
is approximated by a poset. As we shall see in Section \ref{se:4}, $\ca$ is an
infinite-dimensional noncommutative C$^*$-algebra, the poset and its
topology being recoverable from the knowledge of $\ca$. Noncommutative
geometries are built using noncommutative \cstars \cite{Co,VG,CL}.
In this way we
discover the striking result that topologically meaningful finite
approximations to manifolds lead to quantum physics based on
noncommutative geometries.

It bears emphasis that this conclusion
emerges in a natural manner while approximating conventional quantum
theory. Therefore the interest in noncommutative geometry for a physicist
need not depend on unusual space-time topologies like the one used by
Connes and Lott \cite{CL} in building the standard model. Furthermore
these quantum models on posets are of independent interest and
not just as approximations to continuum theories, as they provide us with
a whole class of examples with novel geometries.

In Sections \ref{se:4} and \ref{se:5}, we also discuss many aspects of
quantum physics based on $\ca$,
drawing on known mathematical methods of the noncommutative geometer
and the C$^*$-algebraist.

The \cstars for our posets are as a rule inductive limits \cite{FD}
of finite dimensional algebras, being examples of ``approximately finite
dimensional" algebras \cite{Br,SV}.
Therefore we can approximate $\ca$ by finite dimensional algebras and in
particular by a commutative finite dimensional algebra $\cc(\ca)$. Their
elements can be regarded as continuous ``functions" (or rather, as
sections of a certain bundle) on the poset. They too encode the
topology of the latter. The algebra $\cc(\ca)$
is also strikingly simple, so that it
is relatively easy to build a quantum theory using $\cc(\ca)$. In Section
\ref{se:6},
we describe these approximations and argue also that the approximation
by $\cc(\ca)$ can be obtained from a gauge principle.

Section \ref{se:7} deals with a concrete example having nontrivial
topological
features, namely the poset approximation to a circle. We
establish that global topological effects can be captured by poset
approximations and algebras $\cc(\ca)$ by showing that the
``$\theta $-angle"
for a particle on a circle can also be treated using $\cc(\ca)$.

Section \ref{se:8} concerns the sense in which the algebras
$\ca$ and
$\cc(\ca)$ are continuous ``functions" on their posets. While this
discussion is conceptually important, it was not undertaken earlier to
prevent interruption of the main flow of ideas.
The article concludes with Section \ref{se:9}.

This article is an expanded version of the material covered by the
last lecture of A.P. Balachandran at the XV Autumn School on
``Particle Physics in the Nineties'' [Lisbon, 11-16 October 1993]. The
material covered by the remaining lectures are available elsewhere.
[Chapters 8 and 20 of \cite{BMSS}, and also \cite{Ba}.] It is also an
expanded version of the talk at the International Colloquium on
Modern Quantum Field Theory II [Tata Institute of Fundamental
Research, Bombay, 5-11 January, 1994].

\sxn{The Finite Approximation}\label{se:2}

Let $M$ be a continuous topological space like for example the sphere $S^N$ or
the Euclidean space ${\bf R}^N$. Experiments are never so accurate that
they can detect events associated
with points of $M$, rather they only detect events as occurring
in certain sets $O_\l$. It is therefore
natural to identify any two points $x$, $y$ of $M$ if every set $O_\l$
containing either point contains the other too. Let us assume that the sets
$O_\l$ cover $M$,
\be
   M=\bigcup _\l O_\l~, \label{2.1}
\ee
and write $x\sim y$ if $x$ and $y$ are not separated or distinguished by
$O_\l$ in the sense above:
\be
   x\sim y \mbox{  means  } x\in O_\l \iff y\in O_\l
{}~~~\mbox{for every}~~ O_\l~ . \label{2.2}
\ee
Then $\sim $ is an equivalence relation, and it is reasonable to replace $M$
by $M/ \sim \equiv P(M)$ to reflect the coarseness of observations. It is this
space, obtained by identifying equivalent points [and with the quotient
topology explained later], that will be our
approximation for $M$.

We assume that the number of sets $O_\l$ is finite when $M$ is compact so that
$P(M)$ is an approximation to $M$ by a finite set in this case. When $M$ is
not compact, we assume instead that each point has a neighbourhood
intersected by only finitely many $O_\l$ so that $P(M)$ is a ``finitary"
approximation to $M$ \cite{So}. We also assume that each $O_\l$ is open
\cite{EDM} and that
\be
   \cu =\{O_\l \} \label{2.3}
\ee
is a topology for $M$ \cite{EDM}.
This implies that $O_\l\cup O_\m$ and
$O_\l\cap O_\m$ $\in \cu$ if $O_{\l,\m}\in \cu$. Now experiments can
isolate events in $O_\l\cup O_\m$ and $O_\l\cap O_\m$ if they can do so
in $O_\l$ and $O_\m$ separately, the former by detecting an event in either
$O_\l$ or $O_\m$, and the latter by detecting it in both $O_\l$ and $O_\m$.
The hypothesis that $\cu $ is a topology is thus conceptually consistent.

These assumptions allow us to isolate events in certain sets of the form
$O_\l\setminus [O_\l\cap O_\m]$ which may not be open. This means that there
are in general points in $P(M)$ coming from sets which are not open in $M$.

In the notation we employ, if $P(M)$ has $N$ points, we sometimes denote it
by $P_N(M)$.

Let us illustrate these considerations for a cover of $M=S^1$ by four open
sets as in Fig. 1(a).
In that figure, $O_{1,3}\subset O_2\cap O_4$.
Figure 1(b) shows the corresponding discrete space $P_4(S^1)$, the
points $x_i$ being images of sets in $S^1$. The map $S^1\rightarrow
P_4(S^1)$ is given by
$$
  O_1 \rightarrow x_1, ~ ~ ~ ~ ~
        O_2\setminus [O_2\cap O_4] \rightarrow x_2,
$$
\be
  O_3 \rightarrow x_3, ~ ~ ~ ~ ~
        O_4\setminus [O_2\cap O_4] \rightarrow x_4 ~ .\label{2.3a}
\ee

Now $P(M)$ inherits the quotient topology from $M$ \cite{EDM}.
It is defined as follows.
Let $\Phi $ be the map from $M$ to $P(M)$ obtained by identifying equivalent
points. An example of $\Phi $ is given by (\ref{2.3a}). In the quotient
topology, a set in $P(M)$ is declared to be open if its inverse image for
$\Phi $ is open in $M$. It is the finest topology compatible with the
continuity of $\Phi$. We adopt it hereafter as the topology for $P(M)$.

This topology for $P_4(S^1)$ can be read off from Fig.1, the open sets being
\be
\{x_1\}, ~ ~ \{x_3\}, ~ ~\{x_1,x_2,x_3\}, ~ ~ \{x_1,x_4,x_3\},
{}~ ~ \label{2.4}
\ee
and their unions and intersections (an arbitrary number of the latter being
allowed as $P_4(S^1)$ is finite).

A partial order $\preceq $ \cite{Al,HW,St} can be introduced
in $P(M)$ by declaring that $x\preceq y$ if every open set containing $y$
contains also $x$. It then becomes a \underline{partially ordered set}
or a \underline{poset}. For $P_4(S^1)$, this order reads
\be
x_1\preceq x_2,~~~ x_1\preceq x_4;~~~ x_3\preceq x_2,~~~
x_3\preceq x_4,~~~ \label{2.5}
\ee
where we have omitted writing the relations $x_j\preceq x_j$.

Later, we will write $x\prec y$ to indicate that $x\preceq y$ and $x\neq y$.

In a Hausdorff space \cite{EDM},
there are open sets $O_x$
and $O_y$ containing any
two distinct points $x$ and $y$
such that $O_x\cap O_y=\emptyset $. A
finite Hausdorff space necessarily has the
discrete topology where each point is an open
set. So $P(M)$ is not Hausdorff. But
it is what is called $T_0$ \cite{EDM},
where for any two distinct points, there is an
open set containing at least one of these points and not the other. For
$x_1$ and $x_2$ of $P_4(S^1)$, the open set $\{x_1 \}$ contains $x_1$ and
not $x_2$, but there is no open set containing $x_2$ and not $x_1$.

Any poset can be represented by a Hasse diagram constructed by arranging its
points at different levels and connecting them using the
following rules: 1) If $x\prec y$, then $y$ is higher than $x$. 2) If
$x\prec y$ and there is no $z$ such that $x\prec z\prec y$, then $x$ and $y$
are connected by a line called a link.

In case 2), $y$ is said to cover $x$.

The Hasse diagram for $P_4(S^1)$ is shown in Fig. 2.

The smallest open set $O_x$ containing $x$ consists of all $y$ preceding
$x$ ($y\preceq x$)[so that the closure of the singleton set $\{y\}$
contains $x$].
In the Hasse diagram, it consists of $x$ and all
points we encounter as we travel along links from $x$ to the bottom. In
Fig. 2, this rule gives
$\{x_1,x_2,x_3\}$ as the smallest open set containing $x_2$, just as in
(\ref{2.4}).

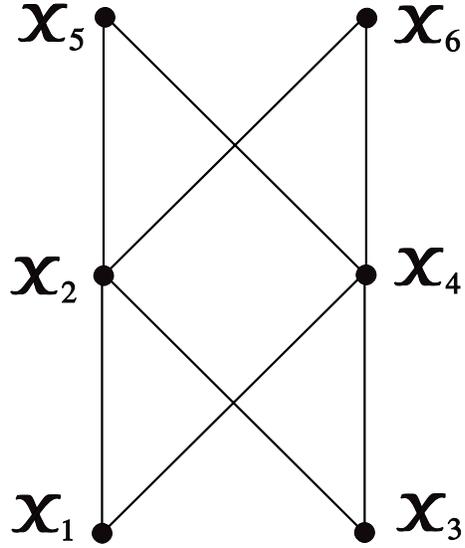
As another example, consider the Hasse diagram of Fig. 3
for a two-sphere
poset $P_6(S^2)$ derived in \cite{So}. Its open sets are generated by
$$
\{x_1\}, ~ ~ ~ \{x_3\}, ~ ~ ~ \{x_1,x_2,x_3\},
{}~ ~ ~ \{x_1,x_4,x_3\}~,
$$
\be
\{x_1,x_2,x_5,x_4,x_3\}, ~ ~ ~ \{x_1,x_2,x_6,x_4,x_3\},
{}~ ~ ~ \label{2.6}
\ee
by taking unions and intersections.

As one more example, Fig. 4 shows a cover of $S^1$ by $2N$ open sets $O_j$ and
the Hasse diagram of its poset $P_{2N}(S^1)$.

Next we define the notion of rank.

A point $x$ of a poset $P$ can be assigned a rank $r(x)$ as follows.
A point of $P$ is regarded as of rank 0 if it converges to no
point, or is a highest point. Let $P_1$ be the poset got from $P$ by
removing all rank zero points and their links. The highest points
of $P_1$ are assigned rank 1. We continue in this way to rank all points.

The rank of a poset is just the maximum rank occurring among the points.
[This definition is commonly used only for ``rankable" posets, a concept
we will not need and will not define in this article]

We conclude this section by illustrating one of the
remarkable properties of a
poset approximation, namely its ability to accurately reproduce
the fundamental group of
the manifold. This we do by indicating that the fundamental group  \cite{fun}
of $P_4(S^1)$ is ${\bf Z}$ \cite{So}.
This group is obtained from
continuous maps of $S^1$ to $P_4(S^1)$, or equivalently, from such maps of
$[0,1]$ to $P_4(S^1)$ with the same value at $0$ and $1$.
Figure 5 shows
maps like this. The maps shown are continuous, the inverse
images of open sets being open \cite{EDM}. The map in Fig. 5(a) can be
deformed to the constant map and has zero winding number. The image in
Fig. 5(b) ``winds once around" $P_4(S^1)$. The map in this figure has
winding number one and is not homotopic to the map in Fig. 5(a). It
leads in a conventional way to the generator of ${\bf Z}$.  Of course
${\bf Z}$ is
also $\p_1(S^1)$.

\sxn{Topology from Quantum Physics}\label{se:3}

In conventional quantum physics, the configuration space is generally a
manifold when the number of degrees of freedom is finite. If $M$ is
this
manifold and $\ch$ the Hilbert space of wave functions, then $\ch$
consists of all square integrable functions on $M$ for a suitable
integration measure. A wave function $\psi $ is only required to be
square integrable. There is no need for $\psi $ or the probability
density $\psi ^*\psi $ to be a continuous function on $M$. Indeed there
are plenty of noncontinuous $\psi $ and $\psi ^*\psi $. Wave functions
of course are not directly observable, but probability densities are,
and the existence of noncontinuous probability densities have
potentially disturbing implications. If all states of the system are
equally available to preparation, which is the case if all self-adjoint
operators are equally observable, then clearly we cannot infer the
topology of $M$ by measurements of probability densities.

It may also be recalled in this connection that any two
infinite-dimensional (separable) Hilbert spaces $\ch_1$ and $\ch_2$
are unitarily related. [Choose an orthonormal basis $\{h_n^{(i)}\}$,
$(n=0,1,2,...)$ for $\ch_i$ $(i=1,2)$. Then a unitary map
$U:\ch_1\rightarrow \ch_2$ from $\ch_1$ to $\ch_2$ is defined by
$Uh_n^{(1)}=h_n^{(2)}$.] They can therefore be identified or thought
of as the same. Hence the Hilbert space of states in itself contains
no information whatsoever about the configuration space.

It seems however that not all self-adjoint operators have equal
status in quantum theory. Instead, there seems to exist a certain class
of privileged observables $\cp \co$ which carry information on the
topology of $M$ and also have a special role in quantum physics.
This set $\cp \co$ contains operators like the Hamiltonian and angular
momentum, and particularly also the set of continuous functions
$\cc(M)$ on $M$, vanishing at infinity if $M$ is noncompact.

In what way is the information on the topology of $M$ encoded in $\cp
\co$? To understand this, recall that an unbounded operator such as a
typical Hamiltonian $H$
cannot be applied on all vectors in $\ch$. Instead, it can
be applied only on vectors in its domain $D(H)$, the latter being dense
in $\ch$ \cite{hil}. In ordinary quantum mechanics, $D(H)$ typically consists
of twice-differentiable functions on $M$ with suitable fall-off
properties at $\infty $ in case $M$ is noncompact. In any event, what
is important to note is that if $\psi $, $\chi \in D(H)$ in elementary
quantum theory, then $\psi ^*\chi \in \cc(M)$. A similar property
holds for the domain $D$ of any unbounded operator in $\cp \co$: If
$\psi $, $\chi \in D$, then $\psi ^*\chi \in \cc(M)$. It is thus in
the nature of these domains that we must seek the topology of $M$.
\footnote{Our point of view about the manner in which topology is
inferred from quantum physics was developed in collaboration with G.
Marmo and A. Simoni.}

We have yet to remark on the special physical status of $\cp \co$
in quantum theory. Let $\ce$ be the intersection of the domains of all
operators in $\cp \co$. Then it seems that the basic physical
properties of the system, and even the nature of $M$, are all inferred
from observations of the privileged observables on states associated with
$\ce$
\footnote{Note in this connection that any observable of
$\cp \co$ restricted to $\ce$ must be essentially self-adjoint \cite{hil}.
This is because if significant observations are all confined to states
given by $\ce$, they must be sufficiently numerous to determine the operators
of $\cp \co$ uniquely.}.

This discussion shows that for a quantum theorist, it is quite important
to understand clearly how $M$ and its topology can be reconstructed from
the algebra $\cc(M)$. Such a reconstruction theorem already exists in
the mathematical literature. It is due to Gel'fand and Naimark
\cite{FD},
and is a basic result in the theory of C$^*$-algebras. Its existence is
reassuring and indicates that we are on the right track in imagining
that it is $\cp \co$ which contains information on $M$ and its topology.

The following however must be noted: The above theorem uses specific
steps to reconstruct $M$ and its topology. It is not so clear that we
actually achieve this reconstruction from physical
experience using the same steps.

We next explain the Gel'fand-Naimark results briefly.

A \cstar $\ca$, commutative or otherwise, is an algebra with a norm
$\parallel \cdot \parallel$ and an antilinear involution * such that
$\parallel a\parallel =\parallel a^*\parallel $, $\parallel
a^*a\parallel =\parallel a^*\parallel ~ \parallel a\parallel $ and
$(ab)^*=b^*a^*$ for $a,b\in \ca $. $\ca$ is also assumed to be complete
in the given norm. Examples of \cstars  are: 1) The algebra of $n\times
n$ matrices $T$ with $T^*=$ the hermitian conjugate of $T$,
and $\parallel T\parallel^2 =$ the largest eigenvalue of $T^* T$;
2) The algebra $\cc(M)$ of continuous functions on $M$,
with * denoting complex conjugation and the norm given by
the supremum norm, $\parallel f\parallel =\sup_{x\in M}|f(x)|$.

In discussing the reconstruction theorem, it is not useful to imagine
that $\cc(M)$ is given as a set of actual functions on $M$, for that
presupposes that $M$ is already known. Rather, it is better to suppose
that we are given a commutative \cstar $\cc$. We must then reconstruct a
topological space $M$ from $\cc $ such that $\cc(M)=\cc$. We would
also like $M$ to be unique up to homeomorphisms.

So we assume that we are given such a $\cc$. Let $M$ denote the space
of [equivalence classes of]
irreducible representations (IRR's), also called the structure space,
of $\cc$. [The trivial IRR given by $\cc\rightarrow \{0\}$ is not included
in $M$.] The \cstar $\cc$ being commutative, every IRR is one-dimensional.
Hence if
$x\in M$ and $f \in \cc$, the image $x(f)$ of $f$ in the IRR defined by
$x$ is a complex number. Writing $x(f)$ as $f(x)$, we can therefore
regard $f$ as a complex-valued function on $M$ with the value $f(x)$ at
$x\in M$. We thus get the interpretation of $\cc$ as {\bf C}-valued functions
on $M$.

We next topologise $M$ by declaring that the set of zeros of each $f\in
\cc$ is a closed set. [This is natural to do since the set of zeros of a
continuous function is closed.] The topology of $M$ is generated by
these closed sets. It is also called the hull kernel or Jacobson topology
\cite{FD}.

Gel'fand and Naimark also show that $\cc=\cc(M)$ and that the
requirement $\cc=\cc(N)$ uniquely fixes the manifold $N$ up
to homeomorphisms. In
this way, we recover the topological space $M$, uniquely up to
homeomorphisms, from the algebra $\cc$
\footnote{
We remark that more refined attributes of $M$ such as a
C$^\infty $-structure can also be recovered using only algebras if more
data are given. For the C$^\infty $-structure, for example, we must also
specify an appropriate subalgebra $\cc ^\infty(M) $ of $\cc(M)$.
The C$^\infty
$-structure on $M$ is then the unique C$^\infty $-structure for which
the elements of $\cc^\infty(M)$ are all the C$^\infty $-functions
\cite{Ma}.}.

We next briefly indicate how we can do quantum theory starting from
$\cc(M)=\cc$.

Elements of $\cc$ are observables, they are not quite wave functions.
The set of all wave functions form a Hilbert space $\ch$. Our first step
in constructing $\ch$, essential for quantum physics, is the
construction of the space $\ce$ which will serve as the common domain of
all the privileged observables.

The simplest choice for $\ce$ is $\cc$ itself
\footnote{Differentiability requirements will in general further
restrict $\ce$. We will as a rule ignore such details in this
article.}.
With this choice, $\cc$
acts on $\ce$, as $\cc$ acts on itself by multiplication. The presence
of this action is important as the privileged observables must act on
$\ce$. Further, for $\psi, ~ \chi \in \ce$, $\psi ^*\chi \in \cc$ exactly
as we want.

Now Gel'fand and Naimark have established that it is possible to
integrate over the structure space $M$ of $\cc$. A scalar product
$(\cdot ,\cdot )$ for elements of $\ce$ can therefore be defined by
choosing an integration measure $d\mu$ on $M$ and setting
\be
        (\psi ,\chi )=\int _M d\mu (x) (\psi ^*\chi )(x).\label{3.1}
\ee
The completion of the space $\ce$ using this scalar product gives the
Hilbert space $\ch$.

The final set-up for quantum theory here is conventional. What is novel
is the shift in emphasis to the algebra $\cc$. It is from this algebra
that we now regard the configuration $M$ and its topology as having been
constructed.

There is of course no reason why $\ce$ should always be $\cc$. Instead
it can consist of sections of a vector bundle over $M$ with a
$\cc$-valued positive definite sesquilinear form $<\cdot ,\cdot >$.
[$<\cdot ,\cdot >$ is positive definite if $<\a,\a>$ is a nonnegative
function for
any $\a\in \ce$ which identically vanishes iff $\a=0$.] The scalar
product is then written as
\be
        (\psi ,\chi )=\int_Md\mu (x)<\psi ,\chi >(x).\label{3.2}
\ee
The completion of $\ce$ using this scalar product as before gives $\ch$.

There is an algebraic construction of nontrivial $\ce$ from $\cc$ which is
meaningful even for noncommutative $\cc$. We will explain this construction
in Section \ref{se:5}.

\sxn{Finite Approximations and Noncommutative Geometry}\label{se:4}

\subsxn{The Algebra $\ca$ for a Poset}\label{se:4.1}

In the preceding section, it has been argued that the algebra $\cc(M)$
plays a basic role in quantum theory on a manifold $M$. It is hence of
interest to enquire about the  algebra $\ca$ replacing $\cc(M)$ when $M$ is
approximated by a poset $P$.

Let us first recall a few definitions and results from operator theory
\cite{hil} before outlining an answer to this question. An operator in a
Hilbert space is said to be of finite rank if the orthogonal complement
of its null space is finite dimensional. It is thus essentially like a
finite dimensional matrix as regards its properties even if the Hilbert
space is infinite dimensional. An operator $k$ in a Hilbert space is
said to be compact if it can be approximated arbitrarily closely in norm
by finite rank operators. Let $\l_1,\l_2,$ ... be the eigenvalues of
$k^* k$ for such a $k$, with $\l_{i+1} \leq \l_i$ and
an eigenvalue of multiplicity $n$
occurring $n$ times in this sequence. [Here and in what follows, $*$
denotes the adjoint for an operator.]
Then $\l_n\rightarrow 0$ as
$n\rightarrow \infty $. It follows that the operator $1\!\!1$ in an infinite
dimensional Hilbert space is not compact.

The set $\ck$ of all compact operators $k$ in a Hilbert space is a \cstar.
It is a two-sided ideal in the \cstar $\cb$ of all bounded operators
\cite{FD,Si}. The *-operation here for both $\ck$ and $\cb$ is hermitian
conjugation.

Note that the sets of finite rank, compact and bounded operators are all
the same in a finite dimensional Hilbert space. All operators in fact
belong to any of these sets in finite dimensions.

The construction of $\ca$ rests on the following result from the
representation theory of $\ck$. The representation of $\ck$ by itself is
irreducible \cite{FD}. It is the \underline{only} IRR of $\ck$ up to
equivalence.
[The
trivial representation where the whole algebra is represented by zero is
ignored here and hereafter.]

The simplest nontrivial poset is $P_2=\{p,q\}$ with $q\prec p$. It is
shown in Fig.6. It is the poset for the interval $[r,s]$ ($r<s$) where the
latter is covered by the open sets $[r,s[$ and $[r,s]$.
The map from subsets of $[r,s]$ to the points of $P_2$ is
\be
\{s\} \rightarrow p~, ~~~[r,s[ \rightarrow q~.\label{4.1.1}~
\ee

The algebra $\ca$ for $P_2$ is
\be
        \ca= {\bf C}1\!\!1 + \ck,\label{4.1}
\ee
the Hilbert space on which
the operators of $\ca$ act being infinite dimensional.

We can see this result from the fact that
$\ca$ has only two IRR's and they are
given by
$$
        p:\l 1\!\!1 +k \rightarrow \l := (\l 1\!\!1 +k)(p)~,
$$
\be
q:\l 1\!\!1 +k \rightarrow \l 1\!\!1 +k := (\l 1\!\!1 +k)(q).
\label{4.1a}
\ee
This remark about IRR's becomes plausible if it is
remembered that $\ck$ has only one IRR.

Now we can use the hull kernel topology for the set $\{p,q\}$.
For this purpose, consider the ``function" $k$. It vanishes at $p$ and
not at $q$, so $p$ is closed. Its complement $q$ is hence open. So of
course is the whole space $\{p,q\}$. The topology of $\{p,q\}$ is thus given
by Fig. 6(a) and is that of the $P_2$ poset just as we want.

We remark here that an
equivalent topology can be defined for finite structure spaces as
follows:
Let $I_x$ be the kernel for the
IRR $x$. It is the (two-sided) ideal mapped to 0 by the IRR $x$. We set
$x\prec y$ if $I_x\subset I_y$ thereby converting the space of IRR's
into a poset. The hull kernel topology is the topology of this poset.

In our case, $I_p=\ck $, $I_q=\{0\}\subset I_p$ and hence $q\prec p$. This
gives Fig. 6(a).

We next consider the $\bigvee$ poset. It can be obtained from the
following open cover of the interval $[0,1]$:
\bea
& & [0,1] = \bigcup_\l \co_\l~: \nonumber \\
& & \co_1 = [0, 2/3[~,~~ \co_2 = ]1/3, 1]~,~~ \co_3 = ]1/3, 2/3[~.
\label{4.2.1}
\eea
The map from subsets of $[0,1]$ to the points of the $\bigvee$ poset
in Fig. 7(a) is given by
\be
[0, 1/3] \rightarrow \a~,~~ ]1/3, 2/3[ \rightarrow \g~,~~
[2/3, 1] \rightarrow \b~.
\label{4.2.2}
\ee

Let us next find the algebra $\ca $ for the $\bigvee$ poset of Fig. 7(a).
The $\bigvee$
 has two
arms 1 and 2. The first step in the construction is to attach an
infinite-dimensional Hilbert space $\ch_i$ to each arm $i$ as in Fig. 7(a).
Let $\cp_i$ be the orthogonal projector on $\ch_i$ in $\ch_1\oplus
\ch_2$ and $\ck_{12}=\{k_{12}\}$ be the set of all compact operators in
$\ch_1\oplus \ch_2$. Then \cite{FD}
\be
        \ca= {\bf C}\cp_1 + {\bf C}\cp_2 + \ck_{12} ~ .\label{4.2}
\ee

The IRR's of $\ca$ defined by the three points of the poset are given by
Fig.7(b). It is easily seen that the hull kernel topology correctly
gives the topology of the $\bigvee$ poset.

The generalization of this construction to any (connected) rank one poset is as
follows. Such a poset is composed of several $\bigvee$'s.
Number the arms and attach an
infinite dimensional Hilbert space $\ch_i$ to each arm $i$
as in Figs. 8(a) and 8(b).
To a $\bigvee$ with arms
$i,i+1$, attach the algebra $\ca_i$ with elements
$\l_i\cp_i+\l_{i+1}\cp_{i+1}+k_{i,i+1}$. Here $\l_{i},\l_{i+1}$ are any two
complex numbers, $\cp_i ,\cp_{i+1}$
are orthogonal projectors on $\ch_i$ and $\ch_{i+1}$ in the Hilbert
space $\ch_i\oplus \ch_{i+1}$ and $k_{i,i+1}$ is any compact operator
in $\ch_i\oplus \ch_{i+1}$. This is as before. But now, for glueing the various
$\bigvee$'s together,
we also impose the
condition that $\l_j=\l_k$ if lines $j$ and $k$ meet at top. The algebra
$\ca$ is
then the direct sum of $\ca_i$'s with this condition:
\be
        \ca= \bigoplus \ca_i ~ ~ ~ \mbox{with $\l_j=\l_k$ if lines $j,k$
meet at top.}\label{4.3}
\ee

Figures 8(a) and (b) also show the values of an element $a=\oplus
[\l_i\cp_i+\l_{i+1}\cp_{i+1}+k_{i,i+1}]$ at the different points of two
typical rank one posets.

There is a systematic construction of $\ca$ for any poset (that is, any
``finite $T_0$ topological space") which generalizes the preceding
constructions for two-level posets. It is explained in Fell and Doran
\cite{FD} and will not be described here. Actually the poset does
not uniquely fix its algebra as there are in general many C$^*$-algebras
with the same poset as structure space \cite{BL}.
[However a Hausdorff structure space
such as a manifold does indeed do so.] The Fell-Doran choice seems to be
the simplest. We will call it $\ca$ and adopt it in this paper.

\subsxn{How to do Quantum Theory using $\ca$}\label{se:4.2}

The noncommutative algebra $\ca$ is an algebra of observables. It
replaces the algebra $\cc(M)$ when $M$ is approximated by a poset. We
must now find the space $\ce$ on which $\ca$ acts, convert $\ce$ into a
pre-Hilbert space and therefrom get the Hilbert space $\ch$ by
completion.

Incidentally, as $\ca$ is supposed to generalize $\cc(M)$, it is also
pertinent to enquire about the sense in which the elements of $\ca$ are
continuous functions on the poset. We postpone the clarification of this
point to Section \ref{se:8}
in order not to interrupt the present discussion.

Now as $\ca$ is noncommutative, it turns out to be important to specify
if $\ca$ acts on $\ce$ from the right or the left. We will take the action
of $\ca$ on $\ce$ to be from the right, thereby making $\ce$ a right
$\ca$-module.

The simplest model for $\ce$ is obtained from $\ca$ itself. As for the
scalar product, note that $(\x^* \h)(x)$ is an operator in a Hilbert space
$\ch_x$ if $\x, ~ \h \in \ca$ and $x \in$ poset. We can hence find a
scalar product $(\cdot ,\cdot )$
by first taking its operator trace $Tr$ on $\ch_x$ and then summing it
over $\ch_x$ with suitable weights $\r_x$:
\be
(\x,\h) = \sum_x \r_x Tr (\x^* \h) (x), ~~~ \r_x \geq 0~. \label{4.4}
\ee

As remarked in Section \ref{se:3}, there is no need for $\ce$ to be $\ca$. It
can be any space with the following properties:
\begin{description}
\item[(1)] It is a right $\ca$-module. So, if $\x \in \ce$ and $a \in
\ca$, then $\x a \in \ce$.

\item[(2)] There is a positive definite ``sesquilinear"
form $<\cdot , \cdot>$ on $\ce$ with values in
$\ca$. That is, if $\x, \h \in \ce$, and $a \in \ca$, then
\item[a)]
\bea
& & <\x, \h> \in \ca~,~~~  <\x, \h> ^\ast =  <\h, \x>~, \nonumber \\
& & <\x, \x> \geq 0 ~~{\rm and}~~  <\x, \x> = 0 \iff \x = 0~.
\eea
\end{description}
\noindent
Here $``<\x, \x > \geq 0"$ means that it can be written as $a^* a$ for
some $a \in \ca$.
\begin{description}
\item[b)]
\be
<\x, \h a> = <\x, \h>a~,~~~ <\x a, \h> = a^* <\x, \h>~.
\ee
\end{description}

The scalar product is then given by
\be
(\x,\h) = \sum_x \r_x Tr  <\x, \h> (x)~. \label{4.5}
\ee

As $\x^* \h(x), <\x, \h>(x), <\x, \h a>(x)$ or $<a\x, \h>(x)$ may not be
of trace class \cite{Si}, there are questions of convergence
associated with (\ref{4.4}) and (\ref{4.5}). We presume that these
traces must be judiciously regularized and modified
(using for example the Dixmier
trace \cite{Co,VG,CL}) or suitable conditions put on
$\ce$ or both. (See also Section \ref{se:5.2}.)
But we will not address
such questions in detail in this article.

When $\ca$ is commutative and has structure space $M$, then an $\ce$ with
the properties described consists of sections of hermitian vector
bundles over $M$. Thus, the above definition of $\ce$ achieves a
generalization of the familiar notion of sections of hermitian
vector bundles to noncommutative geometry.

In the literature
\cite{Co,VG,CL}, a method is available for the
algebraic construction of $\ce$. It works both when $\ca$ is
commutative and noncommutative. In the former case, Serre and Swan
\cite{Co,VG,CL} also
prove that this construction gives (essentially) all $\ce$ of physical
interest, namely all $\ce$ consisting of sections of vector bundles.
It is as follows.
Consider $\ca \otimes {\bf C}^N \equiv \ca^N$ for some integer $N$.
This space consists of $N$-dimensional vectors with coefficients in $\ca$
(that is, with elements of $\ca$ as entries). We can act on it from
the left with $N \times N$ matrices with coefficients in $\ca$.
Let $ e = [e^i_j]$ be such a matrix which is idempotent, $e^2 = e$, and
self-adjoint, $< e \x , \h> = <\x ,e \h >$.
Then, $ e \ca^N$ is an $\ce$, and according to the Serre-Swan theorem,
every
$\ce$ [in the sense above]
is given by this expression for some $N$ and some $e$ for commutative $\ca $.
An $\ce$ of the form $e \ca^N$ is called a ``projective module
of finite type" or a ``finite projective module".

Note that such $\ce$
are right $\ca$-modules. For, if $\x \in e \ca^N$, it
can be written as a vector $(\x^1, \x^2, \cdots, \x^N)$ with $\x^i
\in \ca$ and $e_j^i \x^j = \x^i$. The action of $a \in \ca$ on $\ce$ is
\be
\x \rt \x a = (\x^1 a, \x^2 a, \cdots, \x^N a)~. \label{4.6}
\ee

With this formula for $\ce$, it is readily seen that there are many
choices for $< \cdot~,~\cdot >$~. Thus let $g = [g_{ij}]$, $g_{ij} \in \ca$,
be an $N \times N$ matrix with the following properties: a) $g_{ij}^* =
g_{ji}$;
b) $\x^{i*} g_{ij} \x^j \geq 0$ and $\x^{i*} g_{ij} \x^j = 0 \Leftrightarrow
\x=0$.
Then, if $\h = (\h^1, \h^2, \cdots, \h^N)$ is another vector in $\ce$, we
can set
\be
<\x, \h> = \x^{i*}  g_{ij} \h^j~. \label{4.7}
\ee

In connection with (\ref{4.7}), note that the algebras $\ca$ we consider here
generally have unity. In those cases, the choice $g_{ij} \in {\bf C}$ is a
special case of the condition $g_{ij} \in \ca$. But if $\ca$ has no unity, we
should also allow the choice $g_{ij} \in {\bf C}$ .

The minimum we need for quantum theory is a Laplacian $\D$ and
a potential function $W$, as a Hamiltonian can be constructed from
these ingredients. We now outline how to write $\D$ and $W$.

Let us first look at $\D$, and assume in the first instance that
$\ce = \ca$.

An element $a \in \ca$ defines the operator $\oplus_x a(x)$ on the
Hilbert space $\ch = \oplus_x \ch_x$, the map $a \rt \oplus_x a(x)$ giving a
faithful representation of $\ca$. So let us identify $a$ with
$\oplus_x a(x)$ and $\ca$ with this representation of $\ca$ for the
present.

In noncommutative geometry \cite{Co,VG,CL}, $\D$ is constructed from an
operator $D$ with specific properties on
$\ch$. The operator $D$ must be self-adjoint and the commutator
$[D, a]$ must be bounded for all $a \in \ca$:
\be
D^* = D~,~~~ [D, a]  \in \cb~~~~\mbox{for all}~~ a \in \ca~.
\label{4.7.1}
\ee
Given $D$, we construct the `exterior derivative' of
any $a \in \ca$ by setting
\be
da = [D, a] := [D, \oplus_x a_x] .
\label{4.8}
\ee
Note that $da$ need not be in $\ca$, but it is in $\cb$.

Next we introduce a scalar product on $\cb$ by setting
\be
(\a, \b) = Tr [\a^{*} \b]~,~~~\mbox{for all}~~ \a, \b \in \cb~, \label{4.9}
\ee
the trace being in $\ch$.
[Restricted to $\ca$, it becomes (\ref{4.5}) with $\r_x = 1$. This
choice of $\r_x$ is made for simplicity and can readily dispensed
with. See also the comment after (\ref{4.5})]

Let $p$ be the orthogonal projection operator on $\ca$ for this
scalar product:
\bea
& & p^2 = p^* = p~, \nonumber \\
& & p a = a ~~~\mbox{if}~~~ a \in \ca~, \nonumber \\
& & p \a = 0 ~~~\mbox{if}~~~ (a, \a)=0~~~\mbox{for all}~~~
a \in \ca~. \label{4.9.1}
\eea

The Laplacian $\D$ on $\ca$ is defined using $p$ as follows.

We introduce the adjoint $\d$ of $d$ by writing
\be
(da', da) = (a', \d d a)~, ~~ a', a \in \ca~. \label{4.10}
\ee

At this point we may be tempted to call $\D a  = -\d d a$, following
the definition of the Laplacian on functions in manifold theory.
But that would not be quite correct since (\ref{4.10}) fixes only the
component of $ -\d d a$ in $\ca$. For instance, (\ref{4.10}) suggests
the formula $[D, [D, a]]$ for $\d d a$, but the former may not be in
$\ca$. But $ p \d d a$ is in $\ca$ and is uniquely fixed to be
$p [D, [D, a]]$. Thus we define $\D$ on $\ca$  by
\be
\D a = -p [D, [D, a]]~. \label{4.11}
\ee

As for $W$, it is essentially any element of $\ca$. (There may be
restrictions on $W$ from positivity requirements on the Hamiltonian.)
It acts on a wave function $a$ according to $a \rt aW$,
where $(aW)(x) = a(x)W(x)$~.

A possible Hamiltonian $H$ now is $- \l \D + W~,~ \l > 0$~, while a
Schr\"odinger equation is
\be
i \frac{\pa a}{\pa t} = - \l \D a + a W~. \label{4.12}
\ee

When $\ce$ is a nontrivial projective module of finite type over
$\ca$, it is necessary to introduce a connection and ``lift" $d$
from $\ca$ to an operator $\del$ on $\ce$.
Let us assume that $\ce$ is obtained from the construction
described before (\ref{4.6}).
In that case the definition of $\del$ proceeds as follows.

Because of our assumption, an element $\x \in \ce$ is given by
$\x = (\x^1, \x^2, \cdots, \x^N)$ where $\x^i \in \ca$ and $e^i_j
\x^j = \x^i$. Thus $\ce$ is a subspace of $\ca \otimes {\bf C}^N :=
\ca^N$:
\bea
& & \ce \subseteq \ca^N~, \nonumber \\
& & \ca^N = \{ (a^1, \cdots, a^N)~:~ a^i \in \ca \}~. \label{4.13.1}
\eea
Here we regard $a^i$ as operators on $\ch$. Now $\ca^N$ is a subspace
of $\cb \otimes {\bf C}^N := \cb^N$ where $\cb$ consists of bounded
operators on $\ch$. Thus
\bea
& & \ce \subseteq \ca^N \subseteq \cb^N~, \nonumber \\
& & \cb^N = \{ (\a^1, \cdots, \a^N)~:~ \a^i =
{}~\mbox{bounded operator on}~ \ch \}~. \label{4.13.2}
\eea

Let us extend the scalar product $(\cdot, \cdot)$ on $\ce$ [given by
(\ref{4.7}) and (\ref{4.5})] to $\cb^N$ by setting
\bea
& & < \a, \b > = \a^{i*} g_{ij} \b^j~, \nonumber \\
& & (\a, \b) = Tr <\a, \b> ~~~\mbox{for}~~ \a, \b \in \cb^N~.
\label{4.13.3}
\eea

Next, having fixed $d$ on $\ca$ by a choice of $D$ as in
(\ref{4.7.1}), we define $d$ on $\ce$ by
\be
d \x = (d\x^1, d\x^2, \cdots, d\x^N)~. \label{4.13.4}
\ee
Note that $d\x$ may not be in $\ce$, but it is in $\cb^N$:
\be
d \x \in \cb^N~. \label{4.15.5}
\ee

A possible $\nabla$ for this $d$ is
\be
\nabla \x = e d \x + \r \x~, \label{4.15.6}
\ee
where
\begin{description}
\item[{\rm a)}] $e$ is the matrix introduced earlier,
\item[{\rm b)}] $\r$ is an $N \times N$ matrix with coefficients in $\cb$ :
\be
\r = [\r^i_j]~, ~~\r^i_j \in \cb~, \label{4.15.7}
\ee
\item[{\rm c)}]
\be
\r = e \r e~, \label{4.15.8}
\ee
\item[{\rm and}]
\item[{\rm d)}] $\r$ is anti-hermitian:
\be
< \r \a, \b> + < \a, \r \b> = 0~. \label{4.15.9}
\ee
\end{description}

Note that if $\hat{\r}$ fulfills all conditions but $c)$, then $\r =
e \hat{\r} e$ fulfills $c)$ as well.

Having chosen a $\del$, we try defining  $\del^{*} \del$ using
\be
(\del \x, \del \h) = (\x, \del^{*} \del\h )~, ~~~ \x, \h \in
\ce \label{4.16}
\ee
where $( \cdot, \cdot )$ is defined by (\ref{4.13.3}).
But as in the case of $\ca$, this does not fully determine
$\del^{*} \del \h$. What is fully determined is  $q \del^{*} \del$
where $q$ is the orthogonal projector on $\ce$ for the scalar product
$( \cdot, \cdot )$. We thus define $\D$ on $\ce$ by
\be
\D \eta = - q \del ^{*} \del \eta ~,~~~ \h \in \ce~. \label{4.17}
\ee
Of course, $\D\h \in \ce$~.

A potential $W$ is an element of $\ca$. It acts on $\ce$ according to
the rule ({\bf 1}) following (\ref{4.4}).

A Hamiltonian as before has the form $ - \l \D + W~,~ \l > 0$~.
It gives the Schr\"odinger equation
\be
i \frac{\pa \x}{\pa t} = -
\l \D \x + \x W\, ,~ \x \in \ce ~ . \label{4.15}
\ee

We will not try to find explicit examples for $\D$ here. That task
will be taken up for a simple problem in Section \ref{se:7}.

We will conclude this section by pointing out an interesting property
of states for posets. It does not seem possible to localize a state at the
rank zero points [unless they happen to be isolated points, both open
and closed]. We can see this for example from Fig. 7(b) which shows
that if a probability density vanishes at $\g$, then $\l_i$ (and
$k_{12}$)
are zero and therefore they vanish also at $\a$ and $\b$. It seems possible
to  show
in a similar way that localization in an arbitrary poset is possible
only at open sets. We will briefly interpret this result in the
context of simplicial decompositions in Section \ref{se:6.2}~.

\sxn{Structural Results on $\ca$ Suggest Simple Models
for Quantum States}\label{se:5}

\subsxn{Simple Models for $\ce$}\label{se:5.1}

As mentioned earlier, the \cstars for our posets are generally
of a type called \underline{approximately finite dimensional} (AF)
\cite{Br,SV}. [See also
Section  \ref{se:6.1}.]
They have nice structural properties.
In particular, they have very simple presentations in terms of their
maximal commutative subalgebras, a fact
which can be exploited to
develop relatively transparent models for $\ce$.

Let us start with some definitions \cite{Br,SV}. The commutant $A'$ of a
subalgebra $A$ of $\ca$ consists of all elements of $\ca$ commuting
with all elements of $A$ :
\be
A' = \left \{ x \in \ca : xy = yx~, ~~\forall~ y \in A \right \}~.
\label{5.1}
\ee

A maximal commutative subalgebra $\cc$ of $\ca$ is a commutative
$C^*$-subalgebra of $\ca$ with $\cc ' = \cc$.

The $C^*$-algebras $\ca$ we shall now consider have a unity
$1\!\!1$. We therefore have the concepts of the inverse and unitary
elements for $\ca$: by definition the inverse $a^{-1}$ of $a$ fulfills
$a^{-1}a = a a^{-1} = 1\!\!1$, while $u^* u = u u^* = 1\!\!1$ if $u$
is a unitary element of $\ca$.

Let $\cc$ be a maximal commutative subalgebra of $\ca$ and let $\cu$ be
the normalizer of $\cc$ among the unitary elements of $\ca$:
\be
\cu = \left \{ u \in \ca ~|~ u^* u = 1\!\!1 ~;~ u^* c u \in \cc
{}~~{\rm if}~~ c \in \cc \right \}~.
\label{5.2}
\ee
Then the algebra generated by $\cc$ and $\cu$ is $\ca$ \cite{SV}, the
algebra $\ca$ for a poset being AF. If $M_1, M_2, \dots, $ are subsets
of the $C^*$-algebra $\ca$, and we denote by $<M_1, M_2, \dots, >$ or
by $<~\bigcup_n M_n~>$ the smallest $C^*$-subalgebra of $\ca$
containing $ \bigcup_n M_n$, then the above result can be written as
\be
\ca = < \cc,~ \cu >~.
\label{5.3}
\ee

A property of $\cu$ significant for us is that if $u \in \cu$, then
$u^* \in \cu$, so that $\cu$ is a unitary group.

Let $\Hat{\cc}$ be
the space of IRR's or the structure space of $\cc$. It is a countable
set for the AF algebra $\ca$.

Let $\ell^2(\Hat{\cc})$ be the Hilbert space of square summable
functions on $\Hat{\cc}$:
\be
(g, h) = \sum_x g(x)^* h(x) < \infty ~,~~~\forall g, h \in
\ell^2(\Hat{\cc})~.
\label{5.4}
\ee
We will now see that $\ca$ can be realized as operators on
$\ell^2(\Hat{\cc})$.

For this purpose, note first that each $x \in \Hat{\cc}$
defines an ideal $I_x$ of $\cc$~:
\be
I_x = \{ f \in \cc ~|~ f(x) = 0 \}~.
\label{5.5}
\ee
[Here we regard elements of $\cc$ as functions on $\Hat{\cc}$. Also,
all ideals are two-sided.] Such ideals are called primitive ideals.
They have the following properties for the abelian algebra $\cc$: a)
Every ideal is contained in a primitive ideal, and a primitive ideal is
maximal, that is it is contained in no other ideal; b) A primitive ideal
$I$ uniquely fixes a point $x$ of $\Hat{\cc}$ by the requirement $I_x
=I$. Thus $\Hat{\cc}$ can be identified with the space
Prim$(\Hat{\cc}$) of primitive ideals.

Now if $I_x \in \mbox{Prim}(\Hat{\cc})$ and $c \in \cc$, then $cu^* I_x u =
u^* [ucu^*] I_x u = u^* I_x u$~ since ~$u c u^* \in I_x$.
Similarly  $u^* I_x u c = u^* I_x u$.
Hence $u^* I_x u$ is an ideal.
That being so, there is a primitive ideal $I_y$
containing $u^* I_x u$, $u^* I_x u \subseteq I_y$. Hence $I_x
\subseteq u I_y u^*$. Since $u I_y u^*$ is an ideal too, we conclude
that $I_x = u I_y u^*$ or $u^* I_x u = I_y$.
Calling
\be
y := u^* x = u^{-1} x~,
\label{5.6}
\ee
we thus get an action of $\cu$ on $\Hat{\cc}$.

Next note that $\cc$ in general has unitary
elements and hence $\cu \bigcap \cc \not= \emptyset$. Now $\cu \bigcap
\cc$ is a normal subgroup of $\cu$. We can in fact write $\cu$ as the
semidirect product $\sdp $
of the group $\cu \bigcap \cc$ with a group $U$
isomorphic to $\cu / [\cu \bigcap \cc]$:
\be
\cu = [\cu \bigcap \cc]\sdp U~.
\label{5.7}
\ee
Hence, by (\ref{5.3}),
\be
\ca = < \cc,~ U >~.
\label{5.8}
\ee
This result is of great interest for us.

The group $U$ can be explicitly constructed in cases of our interest.
We will do so below for the two-point and $\bigvee$ posets. The general
result for any rank one poset follows easily therefrom.

It is a striking theorem of \cite{SV} that $\ca$ can be realized as
operators on
$\ell^2(\Hat{\cc})$ using the formul{\ae}~
\bea
& & (h \cdot f)(x) = h(x)f(x)~, \nonumber \\
& & (h \cdot u)(x) = h(u^* x)~, ~~~ \forall f \in \cc~, u \in U~, h \in
\ell^2(\Hat{\cc})~.
\label{5.9}
\eea
We have shown the action as multiplication on the right in order to be
consistent with the convention in Section \ref{se:4.2}. Also the dot
has been introduced in writing this action for a reason which will
immediately become apparent.

This realization of $\ca$ can give us simple models for $\ce$. To see
this, first note that we had previously
used $\ca$ or $e \ca^n$ as
models for $\ce$. But as elements of $\cc$ are functions on
$\Hat{\cc}$ just like $h$, we now discover that they are also
$\ca$-modules in view of (\ref{5.9}), the relation between the dot
product of (\ref{5.9}) and the algebra product (devoid of the dot)
being
\bea
c \cdot f & = & c f \; , \nonumber \\
c \cdot u & = & u c u^{-1} \; . \label{5.9a}
\eea
The verification of (\ref{5.9a}) is easy.

Thus $\cc$ itself can serve as a simple model for $\ce$.

We may be able to go further along this line since certain
finite projective modules over $\cc$ may also serve as $\ce$.
Recall for this purpose that such a module
is $E \cc^N$ where $E$ is
an $N \times N$ matrix with coefficients in $\cc$, which is idempotent
and self-adjoint [$E^2 = E~, E^{*} = E~, ~~{\rm where}~~
E^{* i}_j = E^{j*}_i$]. [Cf. Section \ref{se:4.2}.]
A vector in
this module is $\xi = (\xi^1, \xi^2, \cdots, \xi^N)$ with $\xi^i =
E^i_j a^j~,~ a^j \in \cc$.
Now consider
the action $\xi \rightarrow \xi \cdot u$ where $(\xi \cdot u)^i = u
(E^i_j a^j ) u^{-1}$. The vector $\xi \cdot u$ remains in $E \cc^N$
if
\be
u E^i_j u^{-1} = E^i_j \; , \mbox{ that is }, \; u E u^{-1} = E \; .
\label{5.9b}
\ee
Since $\cc$ anyway acts on $E \cc^N$, we get an action of $\ca$
on $E \cc^N$ when (\ref{5.9b}) is fulfilled. Thus $E \cc^N$ is a
model for $\ce$ when $E$ satisfies (\ref{5.9b}).

The scalar product for $\ell^2(\Hat{\cc})$ written above may not be
the most appropriate one and may require modifications or
regularization as we shall see in Section \ref{se:5.2}.
We only mention that
the problem with (\ref{5.4}) arises because elements of $\ce$ must
belong to $\ell^2(\Hat{\cc})$, a restriction which may be too strong
to give an interesting $\ce$ from $\cc$ or an interesting finite
projective module thereon.

\subsxn{The Two-Point Poset}\label{se:5.2}

We will illustrate the implementation of these ideas for the
two-point, the $\bigvee$ and finally for any rank one poset. That
should be enough to see how to use them for a general poset.

We will treat the two-point poset first. Its algebra is (\ref{4.1}). In
its self-representation $q$, it acts on a Hilbert space $\ch
(=\ch_q)$. Choose an orthonormal basis $h_n~(n = 1, 2, \cdots, )$
for $\ch$ and let $\cp_n$ be the orthogonal projection operator on
${\bf C}h_n$. The maximal commutative subalgebra is then
\be
\cc = < 1\!\!1 ,~ \bigcup_n \cp_n >~.
\label{5.10}
\ee
The structure space of $\cc$ is
\be
\Hat{\cc} = \{ 1, 2, \cdots ; \infty \}~,
\label{5.11}
\ee
where
\bea
a)~~~ n : & & 1\!\!1 \rt 1 := 1\!\!1 (n)~, \nonumber \\
          & & \cp_m \rt \d_{mn} := \cp_m(n)~;
\label{5.12}
\eea
\bea
b)~~~ \infty : & & 1\!\!1 \rt 1 := 1\!\!1 (\infty)~, \nonumber \\
          & & \cp_m \rt 0 := \cp_m(\infty)~.
\label{5.13}
\eea

The topology of $\Hat{\cc}$ is the one given by the one-point
compactification of $\{1, 2, \cdots \}$ by adding $\infty$. A basis
of open sets for this topology is
\bea
& & \{n\}~ ;~ n = 1, 2, \cdots ~~; \nonumber \\
& & \co_k = \{ m ~|~ m \geq k \} \bigcup \{\infty\}~.
\label{5.14}
\eea
A particular consequence of this topology is that the sequence $1, 2,
\cdots, $ converges to $\infty$~.

This topology is identical to the hull kernel
topology \cite{FD}. Thus for instance, the zeros of $\cp_n$ and
$1\!\!1 - \sum_{i=1}^{k-1} \cp_i$ are $\{1, 2, \cdots, \hat{n}, n+1,
\cdots, \infty \}$ and $\{1, 2, \cdots, k-1\}$, respectively, where
the hatted entry is to be omitted. These being closed in the
hull kernel topology, their complements, which are the same as
(\ref{5.14}), are open as asserted above.

The group $U$ is generated by transpositions $u(i,j)~, i \not= j$ of
$h_i$ and $h_j$:
\be
u(i, j) h_i = h_j~, ~~~u(i, j) h_j = h_i~, ~~~u(i, j) h_k =
h_k~~~{\rm if} ~~ k \not= i, j~.
\label{5.15}
\ee

Since the ideals of $n$ and $\infty$ are
\bea
& & I_n = \{\cp_1, \cp_2, \cdots, \Hat{\cp_n} , \cp_{n+1} ,
 \cdots \}~, \nonumber \\
& & I_{\infty} = \{\cp_1, \cp_2, \cdots \}~,
\label{5.16}
\eea
we find,
\bea
& & u(i, j)^{*} I_i u(i, j) = I_j~,~~~
u(i, j)^{*} I_j u(i, j) = I_i~, \nonumber \\
& &  u(i, j)^{*} I_k u(i, j) = I_k ~~~{\rm if}~~~k \not= i,j~,
\nonumber \\
& &  u(i, j)^{*} I_{\infty} u(i, j) = I_{\infty}~,
\label{5.17}
\eea
and
\bea
& & u(i, j) i = j~,~~~ u(i, j) j = i~,~~~ \nonumber \\
& & u(i, j) k = k~~{\rm if}~~~ k \not= i,j~, \nonumber \\
& & u(i, j) \infty = \infty~.
\label{5.18}
\eea

It is worth noting that the representation
(\ref{5.9}) of $\ca$ splits into a direct sum of the IRR's $p, q$
for the two-point poset. The proof
is as follows: $\infty$ being a fixed point for $U$, the functions
supported at
$\infty$ give an $\ca$-invariant one-dimensional subspace. It carries
the IRR $p$ by (\ref{4.1a}) and (\ref{5.13}). And since the orbit of
$n$ under $U$ is $\{1, 2, \cdots \}$, the functions vanishing at
$\infty$ give another invariant subspace. It carries the IRR $q$ by
(\ref{4.1a}) and (\ref{5.12}).

There is a suggestive interpretation of the projection operators
$\cp_n$. [See also the second paper of ref. \cite{Br}.]
The IRR $q$ of $\ca$ corresponds to the open set $[r, s[$
which restricted to $\cc$ splits into the direct sum of the IRR's~
$1, 2, \cdots $~. The IRR of $p$ of $\ca$ corresponds to the point $s$
which restricted to $\cc$ remains IRR. We can think of $1, 2, \cdots $
{}~, as a subdivision of $[r, s[$ into points. Then $\cp_n$ can be
regarded as the restriction to $\Hat{\cc}$ of a smooth function on
$[r, s]$ with the value $1$ in a small neighbourhood of $n$ and the
value zero
at all $m \not= n$ and $\infty$.
In contrast, $1\!\!1$
is the function with value $1$ on the whole interval. Hence it has value
$1$ at all $n$ and $\infty$ as in (\ref{5.12}-\ref{5.13}). This
interpretation is illustrated in Fig 9.

As mentioned previously, there is a certain difficulty in using the
scalar product (\ref{5.4}) for quantum physics.
For the two-point poset, it reads
\be
(g, h) = \sum_n g(n)^* h(n) + g(\infty)^* h(\infty)~,
\label{5.19}
\ee
where $\infty$ is the limiting point of $\{1, 2, \cdots \}$~.
Hence, if $h$ is a continuous function, and $h(\infty) \not=0$,
then Lim$_{n \rt \infty} h(n) = h(\infty) \not= 0$, and $(h, h) =
\infty$. It other words, continuous functions in $\ell^2(\Hat{\cc})$
must vanish at $\infty$. This is in particular true for probability
densities found from $\ce$. It is as though $\infty$ has been deleted
from the configuration space in so far as continuous wave functions
are concerned.

There are
two possible ways out of this difficulty.
a) We can
try regularization and modification of (\ref{5.4}) using some such
tool as the
Dixmier trace \cite{Co,VG,CL}; b) we can try changing the scalar product
for example to $(\cdot~, ~\cdot)'_{\e}~, \e > 0$~, where
\be
(g, h)'_\e = \sum_n \frac{1}{n^{1 + \e}} g(n)^* h(n) + g(\infty)^*
h(\infty)~,
\label{5.20}
\ee
the choice of $\e$ being at our disposal.

There are minor changes in the choice of $u(i,j)$ if this scalar product
is adopted.

\subsxn{The $\bigvee$ Poset and General Rank One Posets}\label{se:5.3}

In the case of the $\bigvee$ poset, there are Hilbert spaces $\ch_1$
and $\ch_2$ for each arm, $\ca$ being the algebra (\ref{4.2}) acting on
$\ch = \ch_1 \oplus \ch_2$. After choosing orthonormal basis
$h_n^{(i)}~, ~i = 1, 2~,~~n = 1, 2, \cdots $~, where the superscript
$i$ indicates that the basis element corresponds to $\ch_i$, and
orthogonal projectors $\cp_n^{(i)}$ on ${\bf C} h_n^{(i)}$, the
algebra $\cc$ can be written as
\be
\cc = < \cp_1, \bigcup_n \cp_n^{(1)} ;  \cp_2, \bigcup_n \cp_n^{(2)} >~.
\label{5.21}
\ee
Here $\cp_i$ are projection operators on $\ch_i$.

The group $U$ as before is generated by transpositions of basis
elements.

The space $\Hat{\cc}$ consists of two sequences $n^{(1)}~, n^{(2)}~
(n = 1, 2, \cdots $)~ and two points $\infty ^{(1)}$, $\infty ^{(2)}$,
with $n^{(i)}$ converging to $\infty^{(i)}$:
\be
\Hat{\cc} = \{ n^{(i)}~, \infty^{(i)}~; i = 1, 2 ~; n = 1, 2, \cdots
\}~. \label{5.22}
\ee
Their meaning is explained by
\bea
& & \cp_i (n^{(j)} ) = \delta_{ij}~, \nonumber \\
& & \cp_m^{(i)} (n^{(j)} ) = \delta_{ij} \delta_{mn}~,
\label{5.23}
\eea
\bea
& & \cp_i (\infty^{(j)} ) = \delta_{ij}~, \nonumber \\
& & \cp_m^{(i)} (\infty^{(j)} ) = 0 ~.
\label{5.24}
\eea

The visual representation of $\Hat{\cc}$ is presented in Fig. 10(a).

The remaining discussion of Section \ref{se:5.2} is readily carried out for
the $\bigvee$ poset as also for a general rank 1 poset.
So we content ourself by
showing the structure of $\Hat{\cc}$ for a $\bigvee\!\!\bigvee$ and a
circle poset in Figs. 10(b),(c).

\sxn{Finite Dimensional and Commutative Approximations}\label{se:6}

\subsxn{Why Approximate $\ca$}\label{se:6.1}

There are several good reasons to try to simplify the preceding approach
based on the algebra $\ca$. The leading reason is a practical one:
$\ca$ is infinite dimensional, and it will greatly ease the pain of
numerical work if it can be approximated by finite dimensional
algebras. A second reason is that our experience with $\ca$ is limited
and there are important technical and physical issues to be resolved
before it can be effectively used as a computational tool. The
divergence of the scalar product
(\ref{5.19}) for constant nonzero functions is an
example of the sort of technical issues encountered for any poset
while a question, both conceptual and physical, relates to the meaning
of the group $U$. For if
for example the points $n$ of Section \ref{se:5.2} have the
interpretation we proposed, then the significance of $U$ gets
unclear. This is because it permutes these points and alters their
order in the interval $[r, s]$ and can not therefore be interpreted in
terms of homeomorphisms of $[r, s]$. This question too appears for any
poset.

Later on in this section, we will propose an interpretation of $U$.
But first, we will describe a sequence of finite dimensional
approximations to $\ca$.
The existence of these finite dimensional approximations is exactly
what characterizes $\ca$ as an approximately finite dimensional, or AF,
algebra.
The leading nontrivial approximation here is
commutative while the succeeding ones are not. The commutative approximation
$\cc(\ca)$ has a suggestive physical interpretation for rank one
posets. Further these approximations correctly capture the topology of the
poset. Being also free of divergences because of their finite
dimensionality, they can thus provide us with excellent models to
initiate practical calculations, and to gain experience and insight
into noncommutative geometry in the quantum domain.

\subsxn{The Approximations}\label{se:6.2}

The algebra $\ca$ for the two point poset of Fig. 6 (a) is ${\bf C}1\!\!1
+ \ck$. Consider the following sequence of $C^*$-algebras of increasing
dimension, the $*$-operation being hermitian conjugation:
\bea
& & \ca_1 = {\bf C}1\!\!1 _{1 \times 1}~, \nonumber \\
& & \ca_2 = {\bf C}1\!\!1 _{2 \times 2} + M(1, {\bf C})~, \nonumber \\
& & \ca_3 = {\bf C}1\!\!1 _{3 \times 3} + M(2, {\bf C})~, \nonumber \\
& & \cdots \nonumber \\
& & \ca_n = {\bf C}1\!\!1 _{n \times n} + M(n-1, {\bf C})~, \nonumber
\\ & & \ldots~~~~~~~~~~~~~.
\label{6.1}
\eea
Here $1\!\!1_{n \times n}$ is the $n \times n$ unit matrix while
$M(n, {\bf C})$ is isomorphic to
the $C^*$-algebra of $n \times n$ complex matrices. A typical
element of $\ca_{n+1}$ is
\be
a_{n+1} = \left[
\begin{array}{cc}
m_{n \times n} &0 \\
0 &\l
\end{array}
\right]~,
\label{6.2}
\ee
where $m_{n \times n}$ is an $n \times n$ complex matrix and
$\l$ is a complex number. Note that the subalgebra $M(n,{\bf C})$
consists of matrices of the form (\ref{6.2}) with the last row
and column zero.

The algebra $\ca_n$ is seen to approach $\ca$ as $n$ becomes bigger
and bigger. We can make this intuitive observation more precise as
thus:
There is an inclusion
\be
F_{n+1,n} ~:~ \ca_n \rt \ca_{n+1}
\label{6.3}
\ee
given by
\be
a_{n} = \left[
\begin{array}{cc}
m_{n-1 \times n-1} & 0 \\
0 &\l
\end{array}
\right]~~ \rt ~~
\left[
\begin{array}{ccc}
m_{n-1 \times n-1} & 0 & 0 \\
0 & \l & 0 \\
0 & 0 & \l
\end{array}
\right]~.
\label{6.4V}
\ee
It is a $*$-homomorphism \cite{FD} since
\be
F_{n+1,n}(a^{*}_{n-1}) =
[F_{n+1,n}(a_{n-1})]^{*}~.
\label{6.5}
\ee
Thus the sequence
\be
\ca_1 \rt \ca_2 \rt \cdots
\label{6.6}
\ee
gives a directed system of $C^*$-algebras. Its inductive limit
is $\ca$ as is readily proved using the definitions in
\cite{FD}.

We must now associate appropriate representations to $\ca_n$ which
will be good approximations to the two-point poset.

The algebra $\ca_1$ is trivial. Let us ignore it. All the remaining
algebras $\ca_n$ have the following two representations:
\begin{description}
\item[a)] The one-dimensional representation $p_n$ with
\be
p_n : a_n \rt \l~.
\label{6.7}
\ee
\item[b)] The defining representation $q_n$ with
\be
q_n : a_n \rt a_n~.
\label{6.8}
\ee
\end{description}

It is clear that these representations approach the representations $p$
and $q$ of $\ca$ as $n \rt \infty$.

The ideal $I_{p_{n}}$ for $p_n$ is
\be
M_{n-1} = \left\{ \left[
\begin{array}{cc}
m_{n-1 \times n-1} & 0 \\
0 &0
\end{array}
\right] \right\}
\label{6.9}
\ee
while the ideal for $q_n$ is
\be
I_{q_n} = \{ 0 \}~.
\label{6.10}
\ee
Hence $I_{q_n} \subset I_{p_n}$ and the hull kernel topology \cite{FD}
gives the two-point poset $\{ p_n, q_n \}$ with $q_n \prec p_n$.
This is shown in Fig. 11 (a). It is exactly the same as the poset in
Fig. 6 (a).

Thus the
preceding two representations of $\ca_{n}$ form a topological space
identical to the poset of $\ca$.

All this suggests that it is good to approximate $\ca$ by $\ca_{n}$ and
regard its representations $p_n$ and $q_n$ as constituting the
configuration space.

In our previous discussions, either involving the algebra $\cc$ or
the algebra $\ca$, we considered only their IRR's. But the
representation $q_n$ of $\ca_{n}$ is not IRR. It has the invariant
subspace
\be
{\bf C} ~ \left( \begin{array}{c} 0 \\ 0 \\ \vdots \\ 0 \\ 1
\end{array} \right)  ~ ~ .\label{6.11}
\ee
In this respect we differ from the previous sections in our treatment
of $\ca_{n}$.

Now consider $\ca _2$. It is a commutative algebra with elements
\be
\left( \begin{array}{cc} \l_1 & 0 \\ 0 & \l_2 \end{array} \right)
\equiv (\l_1,\l_2) ~ , \l_i \in {\bf C} ~ . \label{6.12}
\ee
In this way, we can achieve a commutative simplification of $\ca$.

The cover $[r,s[ \cup [r,s]$ of $[r,s]$ giving the two-point poset is
not a good one in a technical sense, one of
its open sets ($[r,s[$) being entirely contained in the other ($[r,s]$).
There is no good interpretation of this algebra for this reason. We
therefore now turn to the $\bigvee$ poset obtainable from a good cover
of the interval as in Section \ref{se:4.1}.

For the $\bigvee$ poset, the Hilbert space defining $\ca$ is realized as the
direct sum of two Hilbert spaces. Accordingly, we construct the
approximation $\ca _{n+1}$ of $\ca$ by considering
\be
D^{n+1} = {\bf C}^{n+1} \oplus {\bf C}^{n+1}~ . \label{6.13}
\ee
A vector in this space is the direct sum
\be
x \oplus y ~ , ~ x=(x^1 ,x^2 ,\ldots,x^{n+1}) ,
y=(y^1 ,y^2 ~ ,~ \ldots ,y^{n+1}) ~ . \label{6.14}
\ee
We take the scalar product for $D^{n+1}$ to be
\be
(x' \oplus y',x \oplus y) = \sum_i [ x'^{(i)*} x^{(i)} +
y'^{(i)*} y^{(i)}] ~ . \label{6.14a}
\ee
Note also that $D^{n+1}$ has the subspace
\be
E^{n+1} = \{ x \oplus y ~ : ~ x^{n+1}=y^{n+1}=0 \} \label{6.15}
\ee
which is isomorphic to ${\bf C}^n \oplus {\bf C}^n$.

Now let $\cp_1$ be the orthogonal projection on the first ${\bf
C}^{n+1}$ of
$D^{n+1}$ and $\cp_2$ on the second. Let also $L[2n,{\bf C}] =
\{L_{2n}\}$ be the set of all linear operators  on $D^{n+1}$
mapping $E^{n+1}$ to $E^{n+1}$ and its orthogonal complement to $\{0\}$.
Then
\begin{eqnarray}
\ca_{n+1} & = &\{ \l_1 \cp_1 + \l_2 \cp_2 + L_{2n} ~ ; ~ \l_i \in
{{\bf C}} ~ , ~ L_{2n} \in L[2n,{{\bf C}]} \} \nonumber \\
 ~ & = &{{\bf C}} ~ \cp_1 + {{\bf C}} ~ \cp_2 + L[2n,{{\bf C}}] ~ .
\label{6.16}
\end{eqnarray}
The term $L[2n,{{\bf C}}]$ is absent here if $n=0$.

As before, there is a *-homomorphism
\be
F_{n+1,n} : \ca_{n} \rightarrow \ca_{n+1} ~ . \label{6.17}
\ee
Here the image of $L[2(n-1),{\bf C}]\subset \ca_{n}$ is being realized as a
subalgebra of $L(2n,{\bf C})$ in a natural way while the images of $\cp_i$ in
$\ca_n$ are the corresponding $\cp_i$ in $\ca_{n+1}$. We thus have a directed
system of C$^{*}$-algebras. Its inductive limit is $\ca$ \cite{FD}
showing that $\ca_{n}$ approximates $\ca$.

The algebras $\ca_{n}$ have the following three representations:
\begin{eqnarray}
a) ~~ \a_n & : & a_n \rightarrow \l_1 ~ , \nonumber \\
b) ~~ \b_n & : & a_n \rightarrow \l_2 ~ , \nonumber \\
c) ~~ \g_n & : & a_n \rightarrow a_n ~ . \label{6.18}
\end{eqnarray}
Here
\begin{eqnarray}
a_n & = & \l_1 \cp_1 + \l_2 \cp_2 + L_{2(n-1)} ~ ~
{}~ \mbox{ for } n \geq 2 ~,\nonumber \\
a_1 & = & \l_1 \cp_1 + \l_2 \cp_2 ~ . \label{6.19}
\end{eqnarray}

Note that
$\a_n$ and $\b_n$ are abelian IRR's while $\g_n$ is not IRR just like
$q_n$.

Now the kernels of these representations are
\begin{eqnarray}
I_{\a_{n}}& = & {\bf C} \cp_2 + L[2(n-1),{\bf C}] ~ , \nonumber \\
I_{\b_{n}}& = & {\bf C} \cp_1 + L[2(n-1),{\bf C}] ~ , \nonumber \\
I_{\g_{n}}& = & \{ 0 \} ~ . \label{6.20}
\end{eqnarray}

Since
\be
I_{\g_{n}} \subset I_{\a_{n}} \mbox{ and } I_{\g_{n}} \subset I_{\b_{n}} ~ ,
\label{6.21}
\ee
we set
\be
\g_n \prec \a_n ~ , ~ \g_n \prec \b_n ~ . \label{6.22}
\ee
The poset that results is shown in Fig. 12. It is again the $\bigvee$ poset,
suggesting that $\ca_{n}$ and its representations $\a_n,\b_n,\g_n$ are
good approximations for our purposes.\\

Now the C$^{*}$-algebra
\begin{eqnarray}
\ca_{1} & = &\{ a_1 = \l_1 \cp_1 + \l_2 \cp_2 ~ ; ~ \l_i \in
{\bf C}  \} \nonumber \\
 ~& = &{{\bf C}} ~ \cp_1 + {{\bf C}} ~ \cp_2 ~ .
\label{6.23}
\end{eqnarray}
is commutative and its representations $\a_1, \b_1$ and $\g_1$ also capture
the poset topology correctly. Is it possible to interpret $\l_i$?

For this purpose, let us remember that the points of a manifold $M$
are closed, and so are the top or rank zero points of the poset. The
latter somehow approximate the former. Since the values of $a_1$ at
the rank zero points are $\l_1$ and $\l_2$, we can regard $\l_i$ as
the values of a continuous function on $M$ when restricted to this
discrete set. The role of the bottom poset point and the value of
$a_1$ there is to somehow glue the top points together and generate a
nontrivial approximation to the topology of $M$.

We can explain this interpretation further using simplicial
decomposition. Thus the interval $[0,1]$ has a simplicial
decomposition with $[0]$ and $[1]$ as zero-simplices and $[0,1]$ as
the one-simplex. Assuming that experimenters can not resolve two
points if every simplex containing one contains also the other, they
will regard $[0,1]$ to consist of the three points $\a_1 =[0]$, $\b_1
= [1]$ and $\g_1=]0,1[$. There is also a natural map from
$[0,1]$ to these points as in Section \ref{se:2}. Introducing
the quotient topology on these points following that section,
we get back the $\bigvee$ poset. In this approach then,
$\l_1$ and $\l_2$ are the values of a continuous function at the two
extreme points of $[0,1]$ whereas the association of $\l_1 \cp_1 +
\l_2 \cp_2$ with the open interval is necessary to cement the extreme
points together in a topologically correct manner.

[We remark here that the simplicial decomposition of any manifold
yields a poset in the manner just indicated. Recall also the remark
at the  end of Section \ref{se:4.2} that a probability density can not be
localized at rank zero points. This result appears eminently
reasonable in the context of a simplicial decomposition where rank
zero points are points of the manifold. Reasoning like this also
suggests that localization must in general be possible only at the subsets of
the poset representing the open sets of $M$. That seems in fact to be the
case. For, as remarked earlier, localization seems
possible only at the open sets of this poset and the latter
correspond to open sets of $M$.]

We have mentioned in Section \ref{se:4.1}
 that the construction of $\ca$ for
any poset is known \cite{FD}. It is possible to obtain its
approximations $\ca_{n}$ including its commutative simplification
$\cc(\ca)$ therefrom. It is also possible to find the representations
we must use in conjunction with $\ca_{n}$ and $\cc(\ca)$.

We emphasize that the algebra with the minimum number of degrees of
freedom correctly reproducing the poset and its topology seems to be
$\cc(\ca)$.\\

\subsxn{Abelianization from Gauge Invariance} \label{se:6.3}

We have already pointed out that the meaning of the algebra $U
\subset \ca$ is not very clear, even though it is essential to
reproduce the poset as the structure space of $\ca$.

But if its role is just that and nothing more, is it possible to
reduce $\ca$ utilizing $U$ or a suitable subgroup of $U$ in some way
and get the algebra $\cc(\ca)$? The answer seems to be yes in all
interesting cases. We will now show this result and argue also that
this subgroup can be interpreted as a gauge group.

Let us start with the two-point poset. It is an ``uninteresting"
example for us where our method will not work, but it is a convenient
example to illustrate the ideas.

The condition we impose to reduce $\ca$ here is that the observables
must commute with $U$. The commutant $U'$ of $U$ in $\ca$ is just
${\bf C}1\!\!1$. The algebra $\ca$ thus gets reduced to a
commutative algebra, although it is not the algebra we want.

The next example is a ``good" one, it is the example of the $\bigvee$ poset.
The group $U$ here has two commuting subgroups $U^{(1)}$ and $U^{(2)}$.
$U^{(i)}$ is generated by the transpositions $u(k,l;i)$ which permute
only the basis elements $h_{k}^{(i)}$ and $h_{l}^{(i)}$:
\begin{eqnarray}
u(k,l;i) h_{k}^{(i)} & = & h_{l}^{(i)} ~,~u(k,l;i) h_{l}^{(i)} = h_{k}^{(i)}
{}~ ; \nonumber \\
u(k,l;i) h_{m}^{(j)} & = & h_{m}^{(j)} \mbox{ if } m \notin \{k,l\} ~
. \label{6.24}
\end{eqnarray}
These are thus operators acting along each arm of $\bigvee$, but do not act
across the arms of  $\bigvee$. The full group $U$ is generated by $U^{(1)}$
and $U^{(2)}$ and the elements transposing $h_{i}^{(1)}$ and $
h_{i}^{(2)}$.

Let us now require that the observables commute with $U^{(1)}$ and
$U^{(2)}$. They are given by the commutant of $\langle U^{(1)} ,
U^{(2)} \rangle$, the latter being
\be
\langle U^{(1)} , U^{(2)} \rangle ' = {{\bf C}} \cp_1 + {{\bf C}} \cp_2
\label{6.25}
\ee
in the notation of (\ref{4.2}). This algebra being isomorphic to
(\ref{6.23}), we get the result we want.

We can also find the correct representations to use in conjunction
with (\ref{6.25}). They are isomorphic to the IRR's of $\ca$ when
restricted to $\langle U^{(1)} , U^{(2)} \rangle '$. This is an obvious
result.

The procedure for finding the algebra $\cc(\ca)$ and its representations
of interest for a general poset now follows. Associated with each arm
$i$ of a poset, there is a subgroup $U^{(i)}$ of $U$. It permutes the
projections, or equivalently the IRR's [like the $n^{(i)}$ of
(\ref{5.22})] associated with this arm, while having the remaining
projectors, or the IRR's, as fixed points. The algebra $\cc(\ca)$ is
then the commutant of $\langle \bigcup_i U^{(i)} \rangle$ :
\be
\cc(\ca) = \langle \bigcup_i U^{(i)} \rangle ' ~ . \label{6.26}
\ee
The representations of $\cc(\ca)$ of interest are isomorphic to the
restrictions of IRR's of $\ca$ to $\cc(\ca)$.

In gauge theories, observables are required to commute with gauge
transformations. In an analogous manner, we here require the
observables to commute with the transformations generated by
$U^{(i)}$. The group generated by $U^{(i)}$ thus plays the role of
the gauge group in the approach outlined here.

\sxn{A Particle on a Circle}   \label{se:7}

A circle $S^1 = \{e^{i\f}\}$ is an infinitely connected space. It has
the fundamental group \cite{BMSS,fun} ${\bf Z}$. Its universal covering
space \cite{BMSS} is the real line
${\bf R}^1 = \{ x : -\infty < x < \infty \}$.
The fundamental group ${\bf Z}$ acts on ${\bf R}^1$ according to
\be
x \rightarrow x +N ~ , ~ N \in {\bf Z} ~ . \label{7.1}
\ee
The quotient of ${\bf R}^1$ by this action is $S^1$, the projection
map ${\bf R}^1 \rightarrow S^1$ being
\be
\p : {\bf R}^1 \rightarrow S^1 ~
: x \rightarrow e^{ix} ~ . \label{7.2}
\ee

Now the domain of a typical Hamiltonian for a particle on $S^1$ need
not consist of smooth functions on $S^1$. Rather it can be obtained from
functions $\psi_{\theta}$ on ${\bf R}^1$ transforming by an IRR
\be
\r_{\q} : N \rightarrow e^{iN\q} \label{7.3}
\ee
of ${\bf Z}$ according to
\be
\psi_{\q}(x+N) = e^{iN\q} \psi_{\q}(x) ~ . \label{7.4}
\ee
The domain $D_{\q}(H)$ for a typical Hamiltonian $H$ then consists of
these $\psi_{\q}$ restricted to a fundamental domain $0 \leq x \leq
1$ for the action of ${\bf Z}$ and subjected to a differentiability
requirement:
\be
D_{\q}(H) = \{ \psi_{\q} : \psi_{\q}(1) = e^{i\q} \psi_{\q}(0)~;~~
\frac{d\psi_{\q}(1)}{dx} = e^{i\q} \frac{d\psi_{\q}(0)}{dx} \} ~
. \label{7.5}
\ee
[In addition, of course, if $dx$ is the measure on $S^1$ for defining
the scalar product of wave functions, then $H\psi_{\q}$ must be square
integrable for this measure.]
We obtain a distinct quantization, called the $\q$-quantization, for
each choice of $e^{i\q}$.

As we have shown earlier \cite{BBET}, there are similar quantization
possibilities for a circle poset as well. The fundamental group of a
circle poset is ${\bf Z}$. Its universal covering space is the poset
of Fig. 13. Its quotient, for example by the action
\begin{eqnarray}
N   & : & x_j \rightarrow x_{j+3N} ~ , \nonumber \\
x_j & = & a_j \mbox{ or } b_j ~ \mbox{ of Fig. 13 }, ~ N \in {\bf Z}
\label{7.6} \end{eqnarray}
gives the circle poset of Fig. 8(b).\\

In our previous article \cite{BBET}, we argued that the poset analogue
of $\q$-quantization can be obtained from complex
functions $f$ on the poset of Fig. 13 transforming by an IRR of
${\bf Z}$:
\be
f(x_{j +3}) = e^{i\q} f(x_j) ~ .\label{7.7}
\ee
While answers such as the spectrum of a typical Hamiltonian came out
correctly, this approach was nevertheless
affected by a serious defect: continuous complex functions on a
connected poset are constants, so that our wave functions can not be
regarded as continuous.

This defect can now be repaired by using the algebra $\cc(\ca)$
for a circle poset and the corresponding algebra $\bar{\cc}(\ca)$ for
Fig. 13. [See also next section.] If the circle poset is taken to be
the one in Fig. 8(b), then
\be
\cc(\ca) = \{ c=(\l_1 ,\l_2 ) \oplus (\l_2 ,\l_3 ) \oplus (\l_3 ,\l_1 )~
: ~ \l_i \in {{\bf C}} \} ~ , \label{7.8}
\ee
and
\be
\bar{\cc}(\ca) = \{ \bar{c} = \oplus (\m_j , \m_{j+1}) ~:~ \m_k \in
{{\bf C}} \} ~ . \label{7.9}
\ee
Here the values of $c$ at the points of Fig. 8(b) are obtained by
setting $k_{ij} =0$ and replacing $\l_j \cp_j + \l_k \cp_k$ by $(\l_j
,\l_k )$ in that Figure. As for the values of $\bar{c}$ at the points
of Fig. 13, they are given by
\be
\bar{c}(a_j ) = \m_j \; ; \; \bar{c}(b_j ) = (\m_j ,\m_{j+1} ) \; .
\label{7.9a}
\ee

The analogue of (\ref{7.7}) is now the condition
\be
\m_{j+3} = e^{i\q} \m_j \label{7.10}
\ee
while wave functions $\c_{\q}$ are obtained by restricting their
domains to $\{a_j , b_j : 1 \leq j \leq 3 \}$. The result is the
analogue of the ``finite projective module" $\ce$ we encountered
earlier. Let us call it here as $\ce_{\q}$. It is
\begin{eqnarray}
\ce_{\q} = \{ \c_{\q},\c_{\q}',\ldots ~ : ~ \c_{\q} & = & (\m_1 ,\m_2 )
 \oplus (\m_2 ,\m_3 ) \oplus (\m_3 ,e^{i\q}\m_1 ) ~ , \nonumber \\
\c_{\q}'=(\m'_1 ,\m'_2 )&
 \oplus & (\m'_2 ,\m'_3 ) \oplus (\m'_3 ,e^{i\q}\m'_1 ) ~ , ~ \ldots
{}~ ; ~ \m_i , \m'_i \in {\bf C} \} ~ . \label{7.11}
\end{eqnarray}

Here
$\ce_{\q}$ is a $\cc(\ca)$-module, with the action of $c$ on $\c_{\q}$
given by
\be
\c_{\q} c = (\m_1 \l_1,\m_2 \l_2 ) \oplus (\m_2 \l_2,\m_3 \l_3)
\oplus (\m_3 \l_3, e^{i\q} \m_1 \l_1) ~ . \label{7.12}
\ee

It has a sesquilinear form $\langle \cdot,\cdot \rangle$ valued in
$\cc(\ca)$:
\be
\langle \c'_{\q} , \c_{\q} \rangle :=
(\m_1^{'*} \m_1,\m_2^{'*} \m_2 ) \oplus (\m_2^{'*} \m_2,\m_3^{'*} \m_3)
\oplus (\m_3^{'*} \m_3, \m_1^{'*} \m_1) ~ . \label{7.13}
\ee
The associated scalar product is
\be
(\c'_{\q},\c_{\q}) = \sum_{j=1}^3 \{ \langle \c'_{\q} , \c_{\q} \rangle
(a_j) + \langle \c'_{\q} , \c_{\q} \rangle (b_j) \} ~ . \label{7.14}
\ee

We have yet to consider the Laplacian $\D$. We first define it
on $\bar{\cc}(\ca)$ by setting
\be
\D \tilde{c} = \frac{1}{\e^2} \{ \bigoplus (\m_{j-1}+\m_{j+1}-2 \m_j
{}~ ; \m_j+\m_{j+2}-2\m_{j+1}) \} ~ , ~
\label{7.15}
\ee
with $\e$ = a positive constant.
We then restrict $j$ to $1 \leq j \leq 3$ and impose the boundary
condition (\ref{7.10}). This gives us $\D$ on $\ce_{\q}$. The
result is very similar to the one we previously had \cite{BBET}. Notice
especially that the two entries in $\D \tilde{c}$ can be
interpreted in terms of the action of the standard discretized
version of the Laplacian acting on $\m$'s.

The solutions of the eigenvalue problem
\be
\D \c_{\q} = \l \c_{\q} \label{7.16}
\ee
are
\begin{eqnarray}
\l & = & \l_k = \frac{2}{\epsilon ^2}(\cos k -1)~, \nonumber \\
\c_{\q}^{(k)}(a_j) & = & A^{(k)} e^{i kj} + B^{(k)}e^{-i kj} ~ ,\nonumber \\
& & A^{(k)}, B^{(k)}\in {\bf C} \label{7.17}
\end{eqnarray}
where $k = m \frac{2\p}{3} + \frac{\theta}{3}~, m = 1, 2, 3$. [The
expression for $\c_\q^{(k)}(b_j)$ follows from $\c_\q^{(k)}(a_j)$.]
These are exactly our answers in ref. \cite{BBET} but for one
significant difference. In ref. \cite{BBET}, the operator $\D$ did not
mix the values of the wave function at points of rank zero and rank
one, resulting in a double degeneracy of eigenvalues. That unphysical
degeneracy has now been removed because of a better treatment of
continuity properties. The latter prevents us from giving independent values
to continuous probability densities [cf. Section \ref{se:8}] at these two
kinds of points.

It is possible to obtain (\ref{7.15}) using the formula (\ref{4.11}).
But we will not describe that approach here, as the preceding
expression for $\D$ can be easily guessed at.

We have now illustrated several crucial ideas of Section \ref{se:4.2} using
$\ce_{\q}$ and $\cc(\ca)$.

\sxn{Posets as Configuration Spaces and their
Continuous ``Functions''}                             \label{se:8}

In our treatment of quantum physics, there is the underlying notion
that algebras like $\ca$ and $\cc(\ca)$ correspond to the algebra
$\cc(M)$ of continuous functions on a manifold $M$. In the case
of $\cc(M)$, there is a topology on $M$ and on the target space
${\bf C}$, and the continuity of elements of $\cc(M)$ is inferred
therefrom. In contrast, for posets, we have not defined the topology
of the target space until now. We have not therefore yet established
the continuity of $\ca$, $\ca_{n}$ and $\cc(\ca)$ in a sense similar to the
continuity of $\cc(M)$. Our remarks in Section 7 on the lack of
continuity of wave functions in our previous work, and its presence
in the current approach, serve to highlight the significance of this
notion. We now turn to the task of defining the target space topology
and establishing this notion.

Let us consider the algebra $\ca$ to be specific, the treatment of
the other algebras being similar. If $a \in \ca$ and $x$ is a
point of its poset $P$, then $a(x)$ is a
bounded operator on a Hilbert space
$\ch_x$. Let $\ca_x$ denote the set of all these operators $a(x)$ as
$a$ runs over $\ca$.
Then $a$ has values in the space $\bigcup_{x} \ca_{x}$. It
is a section of the bundle
\be
E = \bigcup_{x} \ca _{x} \label{8.1}
\ee
over the poset. It is this $E$ that we must topologize,
and then verify that $\ca$ consists of
continuous maps of $P$ to $E$.

Now there are already topologies $\ct(P)$ and $\ct(\ca)$ for $P$
and $\ca$. The latter is given by the norm $||\cdot ||$, the
$\e$-neighborhood of $a$ being all $b$ such that $||b-a||<\e$.
These
topologies being given, there is not much freedom in the choice of
the topology for $E$. There is in fact a canonical way to find a
topology $\ct(E)$ for $E$ from those of $P$ and $\ca$. Let us now
describe it.

The Cartesian product $P \times \ca$ has the topology $\ct(P) \times
\ct(\ca)$ where a basis of open sets is given by Cartesian products of
open sets in $P$ and $\ca$. There is also a map
\be
\f : P \times \ca \rightarrow E \label{8.2}
\ee
defined by
\be
\f : (x,a) \rightarrow (x,a(x) ) ~ . \label{8.3}
\ee
It is called the evaluation map. The topology $\ct(E)$ is the
finest topology compatible with the continuity of $\f$. A set $U$ in
$E$ is open in this topology if its inverse image $\f^{-1}(U)$ is
open in $P \times \ca$.

Note that the way we define the topology for $E$ here and the way we
defined it for finite spaces approximating manifolds in Section \ref{se:2} are
similar. $\ct(E)$ is
in fact the quotient topology \cite{EDM} for the map
$\f$.

We can now show that any element $a$ in $\ca$ defines a continuous
map from $P$ to $E$ in this topology. For this purpose consider
the subset
\be
S=P \times \{a\} \label{8.4}
\ee
of $P \times \ca$. It inherits the induced topology \cite{EDM} from
$P \times \ca$, its open sets being the Cartesian product of open
sets of $P$ with $\{a\}$. The restriction $\f|_S$ of $\f$ to $S$ is a
continuous map from $S$ to $E$. Now consider the map
$a:P\rightarrow E$ given by $x \rightarrow (x,a(x))$.
If $a^{-1}(U)$ is not open for an open set $U$ in
$E$, then $\f|_{S}^{-1}(U)= a^{-1}(U) \times \{a\}$ is also not
open. That being a contradiction, the continuity of $a$ is established.

We will now verify this result explicitly for the $\bigvee$ poset and the
algebra $\cc(\ca)$, the verification in any other case of interest being
similar.

According to (\ref{6.23}) and Fig. 12(b), a typical element $c_1$ of
$\cc(\ca)$ [or $\ca_1$] for the $\bigvee$ poset is $\l_1 \cp_1 + \l_2
\cp_2$. Its $\e$-neighorhood is defined as stated previously.

Consider the point $(\a_1 ,\l_1 )$ in $E$. Its inverse image is
\be
\{\a_1 \} \times \{ \l_1 \cp_1 + \m_2 \cp_2 \; : \; \m_2 \in
{\bf C} \}~ . \label{8.5}
\ee
Any open set containing (\ref{8.5}) contains the open sets
\be
\{\a_1 , \g_1 \} \times
\{ \l_1 ' \cp_1 + \m_2 ' \cp_2 ~ : ~ ||(\l_1 ' -\l_1 )\cp_1 +
(\m_2 ' - \m_2) \cp_2
||= {\rm sup}(|\l_1 ' -\l_1 |, |\m_2 ' -\m_2 |) < \e \}
\label{8.6}
\ee
for all $\m_2 \in {\bf C}$ and some $\e >0$. Hence it contains the
open set
\be
U_{\e}^{(1)} = \{\a_1 , \g_1 \} \times
\{ \l_1' \cp_1 + \m_2 \cp_2 \; : \; |\l_1 ' -\l_1 | < \e \; ,
\m_2 \in {\bf C} \} \label{8.7}
\ee
for some $\e >0$.

Now consider
\be
\f (U_{\e}^{(1)} ) \subset E \; . \label{8.8}
\ee
Its inverse image is easily seen to be $U_{\e}^{(1)}$. As the latter
is open, we conclude that

a) $\f (U_{\e}^{(1)} )$ is open in $E$;

b) Any open set containing $(\a_1 ,\l_1 )$ contains
$\f (U_{\e}^{(1)} ) $ for some $\e >0$.

But $\f (U_{\e}^{(1)} ) $
also contains $(\g_1 , \l_1 \cp_1 + \m_2 \cp_2)$ for every $\m_2$.
Hence by definition the closure $\bar{ \{
(\g_1 , \l_1 \cp_1 + \m_2 \cp_2) \} }$ of the set $\{
(\g_1 , \l_1 \cp_1 + \m_2 \cp_2) \} $ contains $(\a_1 ,\l_1 )$. In
other words, $(\a_1 ,\l_1 )$ is  a limit point of $\{
(\g_1 , \l_1 \cp_1 + \m_2 \cp_2) \} $. Let us indicate this fact
using an arrow (to suggest convergence) as follows:
\be
(\g_1 , \l_1 \cp_1 + \m_2 \cp_2) \rightarrow (\a_1 ,\l_1 )~, \;
\; \; \forall \m_2 \in {\bf C} \; . \label{8.9}
\ee

We find similarly that

c) Any open set containing $(\b_1 ,\l_2 )$ also contains the open set
$\f (U_{\e}^{(2)} ) $ for some $\e >0$, where
\be
U_{\e}^{(2)} = \{\b_1 , \g_1 \} \times
\{ \m_1 \cp_1 + \l_2 '\cp_2 \; : \; |\l_2 ' -\l_2 | < \e \; ,
\m_1 \in {\bf C} \} \; ; \label{8.10}
\ee

d) $(\b_1 ,\l_2 ) \in \bar{\{  (\g_1 , \m_1 \cp_1 + \l_2 \cp_2) \}
}$ or
\be
(\g_1 , \m_1 \cp_1 + \l_2 \cp_2) \rightarrow (\b_1 ,\l_2 )~, \;
\; \; \forall \m_1 \in {\bf C} \; . \label{8.11}
\ee

Furthermore as one can readily see, if $(\g_1 , \L) \rightarrow (\a_1
,\l_1 )$ and $(\g_1 , \L)\rightarrow (\b_1 , \l_2 )$, then
$\L = \l_1 \cp_1 + \l_2 \cp_2$.

The open sets analogous to $\phi (U_{\e}^{(i)})$ containing
$(\g_1 ,\l_1 \cp_1 + \l_2 \cp_2)$ look
different because $\g_1$ is open in $P$. They are
\be
\f (W_\e ) \label{8.12}
\ee
where
\be
W_\e = \{ \g_1 \} \times
\{ \m_1  \cp_1 + \m_2  \cp_2 ~ : ~
 {\rm sup}(|\m_1  -\l_1 |, |\m_2  -\l_2 |) < \e \} \; .
\label{8.13}
\ee
Any open set containing $(\g_1 , \l_1 \cp_1 + \l_2 \cp_2)$ also
contains $\f (W_\e )$ for some $\e >0$.

We next show the following two results:

a) Every element $c_1=\l_1\cp_1+\l_2\cp_2 $ of $\cc(\ca)$ defines a
continuous map.

b) If a map $P \rightarrow E$ is continuous, it is necessarily a
member of $\cc(\ca)$.

As for (a), $c_1$ is continuous by definition \cite{EDM} if the inverse
image $c_1^{-1}(X)$ of any open set $X \subseteq E$ is open in $P$.
Also by definition, $c_1^{-1}(X)$ is the set in $P$ with an
image under $c_1$ contained in $X$: $c_1 [c^{-1}_1 [X]] \subseteq
X$. Now $c^{-1}_1[\f(U_{\e}^{(1)})]= \{\a_1 ,\g_1 \}$,
$c^{-1}_1[\f(U_{\e}^{(2)})]= \{\b_1 ,\g_1 \}$ and
$c^{-1}_1[\f(W_{\e})] = \{\g_1 \}$ are all
open, proving that $c_1$ is continuous.

As for (b), let $\s : P \rightarrow E$ be a continuous map with
$\s(\a_1)=\l_1$, $\s(\b_1)=\l_2$. As $\s$ is continuous and $\g_1
\rightarrow \a_1$, we have $\s(\g_1) \rightarrow \l_1$. Similarly
$\s(\g_1) \rightarrow \l_2$. Hence by a remark above, $\s(\g_1)
= \l_1 P_1 + \l_2 P_2$ and $\s = \l_1 \cp_1 + \l_2 \cp_2$.

It is worth noting in conclusion that not all maps
$\S : P \rightarrow E$ are continuous.
Thus suppose that $\S(\a_1 ) = \l_1$,
$\S(\b_1 ) = \l_2$ and $\S(\g_1 ) = \m_1 P_1 + \m_2 P_2$ where $\m_1
\neq \l_1$ and/or $\m_2 \neq \l_2$. Then $\S$ is not continuous.

\sxn{Final Remarks}\label{se:9}

In this article, we have reviewed a physically well-motivated approximation
method to continuum physics based on partially ordered sets or posets. These
sets have the power to reproduce important topological features of continuum
physics with striking fidelity, and that too with just a few points. In
addition, as discussed in the previous pages, there is also a remarkable
connection of posets to noncommutative geometry. It is our impression that this
connection is quite deep, and can lead to powerful and
novel schemes for numerical approximations which are also topologically
faithful. They seem in particular to be capable of describing solitons and
the analogues of QCD $\theta $-angles.
Much work of course remains to be
done, but there are already persuasive indications of the fruitfulness of the
ideas sketched in this article for finite quantum physics.

\bigskip
\bigskip
\bigskip

\noindent
{\large \bf Acknowledgements }

This work was supported by the Department of Energy, U.S.A. under
contract number DE-FG02-ER40231. In addition, the work of G.L. was
partially supported by the Italian `Ministero dell' Universit\`a e
della Ricerca Scientifica'.

A.P.B. wishes to thank a) the organisers of the XV Autumn School on
``Particle Physics in the Nineties'' [Lisbon, 11-16 October 1993]
and especially Gustavo Branco, and b) the organisers of the International
Colloquium on Modern Quantum Field Theory II [the Tata Institute of
Fundamental Research, Bombay, 5-11 January, 1994], for their warm hospitality.
Part of the material in this article was presented at these meetings. He is
thankful to them for given him an opprotunity to do so as well. He also
thanks Jos\'{e} Mour\~{a}o for very useful discussions and for drawing
attention to ref. \cite{Ma}.

\end{document}